\documentclass[twocolumn,showpacs,preprintnumbers,nofootinbib]{revtex4}
\usepackage{amsfonts}
\usepackage{graphicx}

\usepackage{graphicx}
\usepackage{dcolumn}
\usepackage{amssymb}
\usepackage{amsmath}
\usepackage{epsfig}

\usepackage[colorlinks=true,linkcolor=red,citecolor=green,urlcolor=cyan,
bookmarks=true,bookmarksopen=true,pdfpagemode=None,pdfstartview=FitH]{hyperref}

\newcommand{\be}{\begin{equation}}
\newcommand{\ee}{\end{equation}}
\newcommand{\bea}{\begin{eqnarray}}
\newcommand{\eea}{\end{eqnarray}}
\def\lapp{\mathrel{\rlap{\raise.5ex\hbox{$<$}}
                    {\lower.5ex\hbox{$\sim$}}}}
\def\gapp{\mathrel{\rlap{\raise.5ex\hbox{$>$}}
                    {\lower.5ex\hbox{$\sim$}}}}

\begin{document}

\title{
\vglue -1.5cm
\vglue -0.3cm
\vskip 0.5cm
\Large \bf
The  2-3 mixing and mass split: 
atmospheric neutrinos and magnetized spectrometers
}
\author{
{Abhijit Samanta$^{a,b}$\thanks{email: \tt 
abhijit.samanta@gmail.com}~~\,and
\vspace*{0.15cm} ~A. Yu. Smirnov$^{b}$\thanks{email:
\tt smirnov@ictp.it}
} \\
{\normalsize\em $^a$Ramakrishna Mission Vivekananda University, Belur Math, Howrah 711 202,
India
\vspace*{0.15cm}}
\\
{\normalsize\em $^{b}$The Abdus Salam International Centre for Theoretical    
Physics} \\
{\normalsize\em Strada Costiera 11, I-34014 Trieste, Italy 
\vspace*{0.15cm}}
}
\thispagestyle{empty}
\vspace{-0.8cm}
\begin{abstract}

We study dependence of the atmospheric $\nu_\mu$ and 
$\bar{\nu}_\mu$ fluxes on the deviations of the 2-3 mixing from maximal, 
$|45^\circ - \theta_{23}|$, on the $\theta_{23}$-octant 
and on the neutrino mass splitting $\Delta m_{32}^2$. 
Analytic expressions for the  $\theta_{23}-$deviation effect 
and the octant asymmetry are derived.
We present conservative estimations of sensitivities of the 
iron (magnetized) calorimeter detectors  (ICAL) to these parameters. 
ICAL can establish the  $\theta_{23}$-deviation at higher than 3$\sigma$  confidence level 
if  $| 45^\circ - \theta_{23}| > 6^{\circ}$ with the exposure of 1 Mton$\cdot$yr. 
Sensitivity to the octant is low for zero or very small 1-3 mixing, 
but it can be substantially enhanced for 
$\theta_{13} > 3^\circ$. ICAL can measure
 the difference of  $\Delta m_{32}^2$ in $\nu$ and $\bar\nu$ channels (the CPT test)
with accuracy   $0.8\times 10^{-4}$ eV$^2$ (3$\sigma$) 
with 1 Mton$\cdot$yr exposure, and the
present MINOS result can be excluded at  $>5\sigma$ confidence level.
We discuss possible ways to further improve sensitivity of the 
magnetized spectrometers.

\end{abstract}
\maketitle

{\large \bf 

}

\section{Introduction}

Determination of the 2-3 mass splitting and leptonic mixing, and in 
particular, the  deviation of $\theta_{23}$  
from the maximal mixing angle, 
\be 
\delta_{23}\equiv 45^\circ -\theta_{23}, 
\ee
is of fundamental importance 
\footnote{$\delta_{23}$ is related to another deviation parameter,
$D_{23} \equiv 1/2 - \sin^2 \theta_{23}$ used in literature as $D_{23} = \sin 2 \delta_{23}$.}. 
Here we use the standard parameterization of the PMNS mixing matrix: 
\be U_{PMNS} = U_{23}(\theta_{23}) \Gamma_\delta U_{13}(\theta_{13}) U_{12}(\theta_{12}),
\label{eq:pmns}\ee 
where $U_{ij}$ is the matrix of rotation in the $ij-$plane, 
and $\Gamma_\delta \equiv {\rm diag}(0, 0, e^{i\delta})$. 
Being maximal or close to maximal, the 2-3 mixing testifies for existence of 
certain underlying symmetry \cite{Mohapatra:2006gs}. 
Comparison of the values of $\delta_{23}$ and $\theta_{13}$  
as well as  the mixing angles in the quarks and lepton sectors  
can shed some light on the origins of 
fermion mass and mixing in general.  

The existing results on $\theta_{23}$ and $\Delta m^2_{23}$ are
summarized  in the Table 1. Note that the global fits 
of oscillation data~\cite{GonzalezGarcia:2010er} (see also \cite{Fogli:2008jx})
show some deviation of the 2-3 mixing from maximal: $\delta_{23}=2-3^\circ$ (1$\sigma$).
Although the data agree well with maximal mixing,  
large deviation, $\delta_{23}=\pm 9^\circ$, 
is still possible.

Concerning the 2-3 mass splitting, the global fit values  are  
in agreement with the results of SuperKamiokande (SK)~\cite{Hosaka:2006zd}
as well as the MINOS measurement in the $\nu$ channel \cite{minos}.

Recently MINOS has reported the values of $\Delta m_{31}^2$ and $\theta_{23}$
in the $\bar\nu$ channel \cite{minos} which differ from those in the $\nu$ channel 
(see tables \ref{tab:th23} and \ref{tab:dmsq}).
If confirmed, this result will testify for an effective (due to existence 
of some new interactions~\cite{Kopp:2010qt}) 
or fundamental CPT violation. The  analysis
of the atmospheric neutrino data does not confirm MINOS result although
the sensitivity of SK  to CPT violation  
is not high since SK sum up effects 
of neutrinos and antineutrinos \cite{Kopp:2010qt}.
Iron calorimeters (ICAL)~\cite{ino} can perform very sensitive search for the CPT violation
and check MINOS result.

New accelerator experiments T2K~\cite{Kato:2008zz} and NO$\nu$A~\cite{Ayres:2004js} 
will improve  precision of  measurements of
$\Delta m_{32}^2$ by factor 2, but their accuracy of measurements of $\theta_{23}$ will
be only slightly better than that of the present global fit (see table \ref{tab:th23}).

There are two aspects of the $\theta_{23}$-measurements:   

\begin{itemize}

\item
determination  of the absolute value of the deviation  
$|\delta_{23}|$, and    

\item
identification of the $\theta_{23}$-octant, {\it i.e.} the sign of $\delta_{23}$, 
or in other words, resolution of the octant degeneracy.  

\end{itemize}

The problem of determination of  $\delta_{23}$ and the octant 
with atmospheric neutrinos has been addressed in a number of publications before 
\cite{Kim:1998bv,Peres:2003wd,Peres:2009xe,GonzalezGarcia:2004cu,
Choubey:2005zy,Indumathi:2006gr,Bandyopadhyay:2007kx}.  
It was realized  \cite{Peres:2003wd,Peres:2009xe,GonzalezGarcia:2004cu} that 
at low energies  oscillation effects on the electron 
neutrino flux are proportional to this deviation, and therefore searches 
for an excess (or suppression) of events in the sub-GeV range 
would testify for $\delta_{23}$.  For water Cherenkov detectors
both  aspects of the 2-3 mixing determinations have been explored in 
\cite{Peres:2009xe,GonzalezGarcia:2004cu}. The study was mainly 
concentrated on effects in the electron neutrino flux.

Magnetized calorimeters are mainly aimed at a detection of the muon neutrinos, 
but they  can distinguish neutrinos and antineutrinos, and this composes their 
main advantage.  These detectors provide better energy and angular resolution of 
the charged  leptons, and  consequently, neutrinos.   
A possibility to disentangle neutrinos and antineutrinos reinforces  
the following features:   
i) The energy and angular resolutions (reconstruction)  are different for 
neutrinos and antineutrinos.  
ii) Sensitivities of the neutrino and antineutrino channels  
to the oscillation parameters 
are substantially different.

Sensitivity of a magnetized calorimeter to the 2-3 mixing and mass splitting  
has been explored in \cite{Choubey:2005zy} and 
\cite{Indumathi:2006gr}.  It has been shown 
that at nonzero value of the 1-3 mixing the octant discrimination 
is more feasible with the magnetized detector 
than with the water Cherenkov detector  
since the former can directly measure the  matter effect \cite{Choubey:2005zy}. 
In these studies, however, various simplifications have been made  
which do not allow for realistic estimations of  potential of the experiments. 
In the  analysis~\cite{Samanta:2008af} a possibility of the octant discrimination 
has been evaluated  for two benchmark values of $\theta_{23}$ and 
relatively high $\theta_{13}=7.5^\circ$. 

Here we present  a comprehensive study of  sensitivities of the ICAL to the 
parameters of 2-3 sector. We  assume that by the time of 
operation of this detector certain information about  
$\theta_{13}$ will be obtained.

The paper is organized as follows.
In sec. II  we study dependence of the $\nu_\mu$ and 
$\bar{\nu}_\mu$ fluxes on parameters of the 2-3 sector: 
$|\delta_{23}|,$ 
octant and $\Delta m_{23}^2$. In sec. III we describe details of our 
statistical analysis.  We evaluate physics potential of the magnetized calorimeter  
to measure these parameters in sec. IV. 
In sec. V we consider further improvements of 
sensitivities of the magnetized calorimeters.
Conclusions are given in sec. VI.

\section{Dependence of the atmospheric neutrino fluxes on 2-3 mixing}
\label{s:osc}

The original atmospheric neutrino flux contains both the muon, $F_\mu^0$, and 
electron, $F_e^0$, neutrino components, so that the muon neutrino flux at a detector  equals 
\bea
F_\mu & = & F_\mu^0 P_{\nu_\mu \rightarrow \nu_\mu} 
+  F_e^0 P_{\nu_e \rightarrow \nu_\mu} 
\nonumber\\
& = & 
F_\mu^0 \left[ P_{\nu_\mu \rightarrow \nu_\mu} + \frac{1}{r}
P_{\nu_e \rightarrow \nu_\mu} \right].
\label{eq:muprob}
\eea
Here the flavor ratio
\be
r(E, \theta_Z) \equiv \frac{F_{\mu}^0 (E, \theta_Z)}{F_e^0(E, 
\theta_Z)}\, 
\label{ratior}
\ee 
is function of the neutrino energy $E$ and 
zenith angle $\theta_Z$. 

For the standard parameterization of the mixing matrix 
dependence of the oscillation probabilities on the 2-3 mixing 
$\theta_{23}$ and CP-phase $\delta$ is {\it explicit} for an arbitrary 
density profile. This follows from the order of rotation in eq.~(\ref{eq:pmns}) 
and the fact that the matrix of matter potentials  
has the form  $V =  diag\{ V_e, 0, 0\}$ 
in the flavor basis, {\it i.e.} it is invariant under the 2-3 rotations. 
Indeed, the neutrino evolution can  
be considered in the propagation basis,  
$\tilde{\nu} \equiv (\nu_e, \tilde{\nu}_{2}, \tilde{\nu}_{3})$,
defined via the following relation with the flavor basis:  
$\nu_f \equiv U_{23} \Gamma_\delta \tilde{\nu}$.   
Consequently,  $\tilde{\nu} = U_{13} U_{12} \nu_{mass}$, where  
$ \nu_{mass} \equiv  (\nu_1, \nu_2, \nu_3) $ is the basis of 
mass eigenstates. In the propagation basis  the Hamiltonian, 
and therefore the amplitudes of transitions 
depend on $\theta_{13}, \theta_{12}, V_e$, and mass squared 
differences: 
\be
A_{\alpha \beta} =  A_{\nu_\alpha \rightarrow \nu_\beta}(\theta_{13}, \theta_{12}, V_e), 
~~~~\alpha,  \beta = e, \tilde{2}, \tilde{3},
\label{eq:a-prop}
\ee
and they do not depend on $\theta_{23}$ and $\delta$. 
In the flavor basis,  dependence of the amplitudes $A_f$ 
on these parameters appears via projections of the matrix of amplitudes 
(\ref{eq:a-prop}) from the propagation 
basis to the flavor basis: 
\be
\hat{A}_f  =  
U_{23} \Gamma_\delta \hat{A} \Gamma_{-\delta} U_{23}^T ,  
\label{eq:flav}
\ee
where $\hat{A}^f $ is the matrix of 
amplitudes in the flavor basis.   

In terms of the amplitudes $A_{\alpha\beta}$ using 
eq.~(\ref{eq:flav}) one can 
rewrite the expression for the muon neutrino flux, eq.~(\ref{eq:muprob}),  
in the following form~\cite{Akhmedov:2008qt}: 
\bea
\frac{F_{\mu}}{F_{\mu}^0} & \approx & 1 - \sin^2 2\theta_{23}\sin^2 
\frac{\phi}{2}
\nonumber\\     
& - & \frac{1}{2} \sin^2 2\theta_{23} \cos \phi \left[1 -  {\rm Re} 
(A_{\tilde{2}\tilde{2}}^*A_{\tilde{3}\tilde{3}}) \right] 
\nonumber\\
& - & \left(s_{23}^4 - \frac{s_{23}^2}{r}\right) \tilde{P}_A 
- \left(c_{23}^4 - \frac{c_{23}^2}{r}\right) \tilde{P}_S
\nonumber\\
& - & \sin 2\theta_{23} P_\delta,
\label{eq:flrat}
\eea
where 
\be
\tilde{P}_A \equiv |A_{e\tilde{3}}|^2 ~~{\rm and}~~ \tilde{P}_S \equiv |A_{e\tilde{2}}|^2 
\ee
are the probabilities of transitions 
$\nu_e \rightarrow \tilde{\nu}_{2}$ and $\nu_e \rightarrow \tilde{\nu}_{3}$ 
correspondingly,  
and $P_\delta$ is the function which depends on the CP-violation phase $\delta$.
In eq.~(\ref{eq:flrat}) $\phi$ is the oscillation phase due to the 2-3 
mass splitting: $\phi =  \Delta m_{32}^2x/2E$,  
$c_{23} \equiv \cos^2 \theta_{23}$, {\it etc.}. 
Notice that $\tilde{P}_S$, $\tilde{P}_A$,  $P_\delta$ and $\phi$ \
do not depend on $\theta_{23}$. 

In eq.~(\ref{eq:flrat}) the first two terms are due to vacuum oscillations 
driven by the 2-3 mixing and mass splitting; the second line describes  
interference of the 2-3 and 1-2 modes of oscillations. 
The product of amplitudes in this term 
can be approximated as 
\be
|A_{\tilde{2} \tilde{2}} A_{\tilde{3} \tilde{3}} | 
\approx  \sqrt{(1 - \tilde{P}_A) (1 - \tilde{P}_S)}.   
\ee
The terms in the third line  describe effects of oscillations due to the 1-2 and 1-3 mixing.
The last term in (\ref{eq:flrat}) describes  the CP-violation.  
The leading  (second) term as well as the  interference and CP-violating terms 
are symmetric with respect to change of sign 
of the deviation: 
$\delta_{23} \rightarrow - \delta_{23}$. 
The octant symmetry (degeneracy) is broken by the terms in the third line  
of eq. (\ref{eq:flrat}); these terms    
vanish for the maximal 2-3 mixing and $r = 2$. 

For antineutrinos one obtains the same formula  as in eq.~(\ref{eq:flrat}) 
with substitution $\tilde{P}_S \rightarrow \bar{\tilde{P}}_S,$ 
$\tilde{P}_A \rightarrow \bar{\tilde{P}}_A$ and $r \rightarrow \bar{r}$. 

The octant effect can be characterized by the octant asymmetry defined as
\be
\frac{\Delta^{oct} F_\mu}{F^0_\mu}  \equiv \frac{1}{F^0_\mu}\left[ F_\mu(45^{\circ} + \delta_{23}) -
F_\mu(45^{\circ} - \delta_{23})\right].   
\label{eq:dev}
\ee
For such symmetric deviations from the maximal 2-3 mixing one has 
\bea
\Delta (\sin^2 2\theta_{23}) &  = & 0, 
\nonumber\\ ~~~ 
\Delta (\cos^2 \theta_{23}) & = & \Delta (\cos^4 \theta_{23})  = - \sin 2\delta_{23}.  
\eea
Then according to eq.~(\ref{eq:flrat})  the octant asymmetry equals 
\be
\frac{\Delta^{oct} F_\mu}{F_\mu^0} =
\sin 2 \delta_{23} \, \left(1 - \frac{1}{r} \right)\, (\tilde{P}_A - \tilde{P}_S).  
\label{eq:deltaf}
\ee
Notice that $\tilde{P}_S$ and $\tilde{P}_A$ enter the asymmetry with opposite 
signs, and therefore partly cancel each other. 

To get an idea about dependences of the probabilities on the neutrino parameters 
one can use  expressions for the amplitudes in medium with constant density: 
\bea
A_{e\tilde{2}} & = & c_{13}^m \sin 2 \theta_{12}^m \sin \phi_{21}^m~, 
\nonumber\\
A_{e\tilde{3}} & = & \sin 2\theta_{13}^m  \left(\sin \phi_{32}^m e^{-i \phi_{31}^m} 
+ \cos^2 \theta_{12}^m \sin \phi_{21}^m
\right).
\eea
The probability $\tilde{P}_S \propto \sin^2 2\theta^m_{12}$ 
decreases with energy, whereas $\tilde{P}_A \propto \sin^2 2\theta^m_{13}$ 
increases being resonantly enhanced in the neutrino channel 
(for the normal mass hierarchy) at high energies. 
Here sensitivity of ICAL to the sign of muon charge will play important role. 
The two probabilities become comparable at 1 GeV 
for  $\sin^2 \theta_{13} \sim 0.01$.

In the limit of zero 1-3 mixing 
$\tilde{P}_A = 0$, $\tilde{P}_S \rightarrow {P}_S$,  
and one finds from eq.~(\ref{eq:flrat}) 
\bea 
\frac{F_{\mu}}{F_{\mu}^0}  \approx  
1 - \sin^2 2\theta_{23} \sin^2\frac{\phi}{2} 
- \frac{1}{2} \sin^2 2\theta_{23} \cos \phi \times \nonumber\\
\times \left(1 -  \sqrt{1 - P_S}\right) 
- \left(c_{23}^4 - \frac{c_{23}^2}{r}\right) P_S,  
\label{eq:mumu} 
\eea
where 
${P}_S = \tilde{P}_S(\theta_{13} =  0)$,
which is the $2\nu$ probability of 
oscillations driven by $\Delta m^2_{21}$ and $\theta_{12}$. 
For the octant asymmetry we obtain
\be
\frac{\Delta^{oct} F_\mu}{F_\mu^0} = 
- \sin 2 \delta_{23} \, \left(1 - \frac{1}{r} \right) \, P_S.  
\label{eq:deltaf}
\ee
At low energies: $r \approx 2,$ and therefore the asymmetry  equals 
$\sim 1/2 \, \sin 2 \delta_{23} \, P_S$. 

\begin{figure*}[htb]
\includegraphics[width=8.0cm,angle=0]{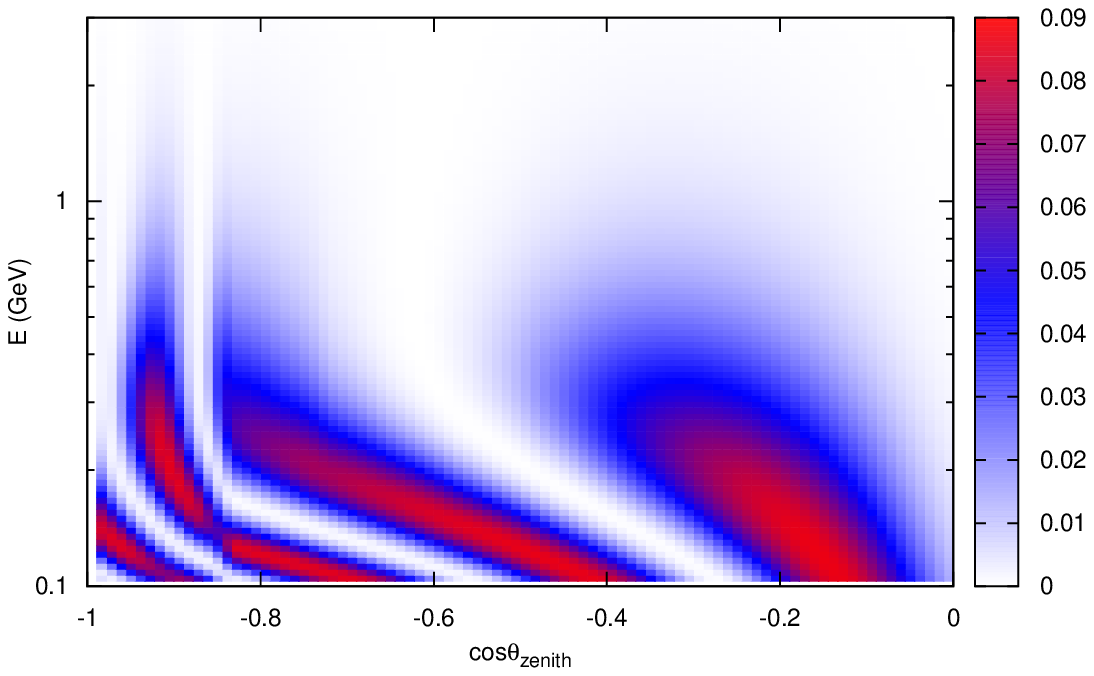}
\includegraphics[width=8.0cm,angle=0]{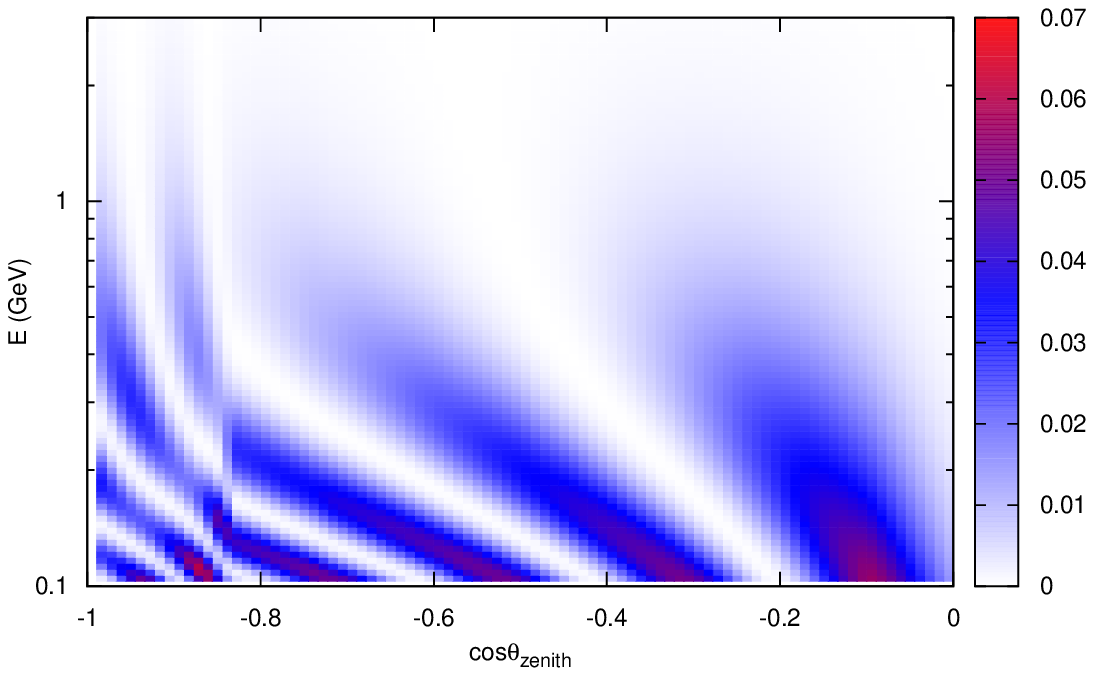}
\caption{\sf \small The oscillograms for the  octant asymmetry: 
The contours of equal asymmetry  $\Delta^{oct} F_\mu / F_\mu^0$ 
for neutrino (left) and antineutrino (right) 
with $\delta_{23} = 5^{\circ}$ at $\theta_{13}=0.$
The other parameters are set at their best-fit values. 
}
\label{f:dhoct}
\end{figure*}

In fig.~\ref{f:dhoct} we show the oscillograms 
for the octant asymmetry, that is, the  lines 
of equal asymmetry $\Delta^{oct} F_\mu /F_\mu^0$ in the $E-\cos\theta_Z$ plane
in neutrino and antineutrino channels. 
According to eq.~(\ref{eq:deltaf}) these oscillograms coincide 
with the oscillograms for $P_S$ 
up to coefficient which weakly depends on $E$ and $\theta_Z$ 
at $E<1$ GeV.
We use  $\delta_{23} = 5^{\circ}$, $\theta_{13}=0,$ and 
other parameters are set at their best-fit values.
The asymmetry increases with decrease of the neutrino energy.
Maximal asymmetry is achieved in the 1-2 resonance ($E \sim 0.1$ GeV): 
$\Delta^{oct} F_\mu/F_\mu^0  \approx 0.087$. 
For realistic threshold $E_{th} = 0.3$ GeV and $\delta_{23} = 5^{\circ}$ 
the averaged asymmetry is about (2 - 3)\% and for 
$E_{th} =  0.8$ GeV it  is below 1\%.  

Notice that the octant asymmetry 
of $\nu_e-$flux is about 4 times larger 
than the $\nu_\mu-$flux asymmetry: 
\be
\frac{\Delta^{oct} F_e}{F_e^0} =
- \sin 2 \delta_{23} \, r \,  P_S.
\label{eq:deltafe}
\ee
Here, however, the original $\nu_e-$flux is 2 times smaller.  
Furthermore, detection of muons provide better 
energy and direction resolutions. 

Since $P_S$ and $\bar{P}_S$ are of the same order,  
separation of the neutrino and antineutrino 
signals has no sense here. 

Let us consider variations of the $\nu_\mu$-flux due to 
deviation of the 2-3 mixing from maximal. 
According to (\ref{eq:flrat}) the relative change of the flux,  
which we will call the $\theta_{23}-$deviation effect,
equals
\bea
\frac{\Delta^{dev} F_\mu}{F_\mu^0} & \equiv & 
\frac{F_\mu(45^{\circ}) - F_\mu(45^{\circ} - \delta_{23})}{F_\mu^0} \approx
\nonumber\\
&  - &  \Delta (\sin^2 2\theta_{23}) \sin^2 \frac{\phi}{2}  
\cong - \frac{1}{2} \sin^2 2\delta_{23}, 
\label{eq:deltadev}
\eea
where in the last equality we have averaged the oscillations due to large mass 
splitting. 

Ratio of the octant asymmetry and the 2-3 deviation effect equals 
\be
\frac{\Delta^{oct} F_\mu}{\Delta^{dev} F_\mu} = 
- \frac{1}{\sin 2\delta_{23}} \left(1 - \frac{1}{r}\right) 
(\langle \tilde{P_A} \rangle - \langle \tilde{P_S} \rangle ),    
\ee
where $\langle \tilde{P_S} \rangle$  and  $\langle \tilde{P_A} \rangle$ are  
the probabilities averaged over the experimental  
$E-\cos\theta_Z$ ranges. 
Although $\Delta^{dev} F_\mu^0$  is proportional to square of the deviation parameter, 
for not very small 
$\delta_{23}$ ($> 5^{\circ}$) and $\theta_{13}=0$, the integral 
effect of the deviation from maximal mixing is stronger 
than the effect of  octant. The reason is that the deviation 
effect does not change with energy, whereas 
the octant asymmetry being proportional to  $P_S$ quickly decreases with $E$
($\langle\tilde{P_S} \rangle \ll \langle \tilde{P_A}\rangle).$
For zero 1-3 mixing and  $\delta_{23} = 5^{\circ}$ we have $\Delta^{dev} F_\mu/F_\mu^0 = 0.015$,  
and $\Delta^{oct} F_\mu/F_\mu^0 =   0.087 \langle P_S \rangle$.

For non-zero 1-3 mixing at high energies the ratio of the flux differences is 
\be
\frac{\Delta^{oct} F_\mu}{\Delta^{dev} F_\mu} \cong  - \left( 1 - \frac{1}{r}\right) 
\frac{\langle \tilde{P}_A \rangle}{\sin 2\delta_{23}}, 
\ee
and since $\tilde{P}_A$ does not decrease with energy the ratio 
is not small in contrast to the previous case.  
 
In our studies of sensitivities to the parameters of 2-3 sector 
we obtain the  oscillation probabilities solving numerically 
full three flavor evolution equation and  using 
the Preliminary Reference Earth Model (PREM)  
\cite{Dziewonski:1981xy} for the density profile of the Earth.
We will use the consideration presented in this section 
for interpretation of the numerical results.

\section{The $\chi^2$ analysis for ICAL}\label{s:chi}

To evaluate  physics potential of the atmospheric neutrino studies with 
a magnetized ICAL detector 
we generated the atmospheric neutrino events  and
considered the muon energy and direction (directly measurable quantities) 
using event generator NUANCE-v3 \cite{Casper:2002sd}. 
The GEANT \cite{geant} simulation of ICAL detector shows  that 
the energy and angular resolutions of muons are very
high  and the corresponding uncertainties  are negligible compared to 
differences between the  angles as well as energies of 
the neutrino and muon in the scattering (production) process.

$\chi^2$ is calculated according to the Poisson distribution.   
The term due to contribution of
prior information on the oscillation parameters measured by other
experiments is not added to $\chi^2$ here for conservative estimation; 
the effect of priors will be considered in sec. V. 
The data have been binned in cells of equal size  in 
the  ($\log_{10} E$ - $L^{0.4}$) plane, where $L=2 R \cos\theta_Z$ 
is the length of neutrino trajectory. Choice of this binning is motivated by 
pattern of the oscillation probability  
$P(\nu_\mu \rightarrow \nu_\mu)$ in the 
$L-E$ plane \cite{Samanta:2008ag}. 
The distance between two consecutive 
oscillation peaks in this plane increases (decreases) 
as one goes to lower $L$ ($E$) values for a given $E$ ($L$).
The binning of $L$ has been optimized to get better sensitivity
to the oscillation parameters.  
To maintain $\chi^2/d.o.f. \approx 1$ in Monte Carlo simulation study, 
number of events should be  $>4$ per cell \cite{Samanta:2008af}.
If the number is smaller than 4 (which happens in the high energy bins), 
we combine results from the nearest cells.

For each set of oscillation parameters we integrate the 
atmospheric neutrino flux at the detector over the energy and zenith angle  
folding it with the cross-section, the  exposure time, the target mass, the efficiency 
of detection and the two dimensional energy-angle correlated 
resolution functions to obtain data for the $\chi^2$ analysis.
We use the charge current cross section of NUANCE-v3 \cite{Casper:2002sd} and the
neutrino flux of  the 3-dimensional scheme \cite{Honda:2006qj}.

The systematic uncertainties of the atmospheric neutrino flux 
are crucial for determination  of the oscillation parameters. 
We have  divided them into two categories:
(i) the overall flux normalization uncertainties which are independent of
the neutrino energy and zenith angle, and
(ii) the spectral tilt uncertainties which depend on $E$ and $\theta_Z$. 
The flux with uncertainties included  can be written as 
\bea
&&\Phi (E,\theta_Z)  =  
\Phi_0(E) \left[ 1 + \delta_E \log_{10} \frac{E}{E_0} \right] \nonumber\\
     && \times \left[ 1 + \delta_Z (|\cos\theta_Z|-0.5) \right]  \times 
\left [1+\delta_{f_N}\right]. 
\label{e:uncer}
\eea
For $E < 1$ GeV we take the energy-dependent uncertainty: $\delta_E=15\%$ and $E_0=1$ GeV, 
and for $E>10$  GeV  correspondingly, $\delta_E=5\%$ and  $E_0=10$ GeV.
The uncertainty is $\delta_E \sim 7\%$ in the range  $E = 1-10 $ GeV. 
The overall  flux uncertainty as  function of the zenith angle  
is parameterized by $\delta_Z$.  
According to  \cite{Honda:2006qj} we use  $\delta_Z=4\%$ which leads to 2\% 
vertical/horizontal flux uncertainty.  
We take for  the overall flux normalization uncertainty $\delta_{f_N}=10\%$  
and for the neutrino cross-section uncertainty: $\delta_{\sigma}=10\%$.

In our  $\chi^2$ analysis the numbers of events have been computed for 
the theoretical (fit) values  and experimental (true) values of parameters 
in the same way by migrating the number of events from the neutrino 
to muon energy and zenith angle bins. The resolution  
functions have been taken from  the previous work \cite{Samanta:2009qw}.

In studies of sensitivity to  the 2-3 mixing we marginalize  $\chi^2$ over 
$\Delta m_{32}^2, ~\theta_{13}$, $\delta_{CP}$
for  $\nu$'s and $\bar\nu$'s separately.  
Then we find the total $\chi^2$ as $\chi^2=\chi^2_\nu+\chi^2_{\bar\nu}$.
We have chosen the following benchmark values of the  neutrino parameters: 
$\Delta m_{32}^2 = 2.5 \times 10^{-3}$ eV$^2$, $\delta_{CP}=0$,
$\Delta m_{21}^2 = 7.9  \times 10^{-5}$ eV$^2$ and $\theta_{12}=34.24^\circ$.
In marginalization we use, first,  flat distributions of values 
of the parameters in the following ranges: 
$\Delta m_{32}^2= (2.3 - 2.7) \times 10^{-3}$eV$^2$,
$\theta_{23}= 36^\circ - 54^\circ $ and  $\theta_{13} =  0^\circ - 10.5^\circ$.
The range of $\theta_{13}$ is changed for some particular analyses. 
(The non-flat distributions of parameters will be considered in sec. V.)

The parameters $\Delta m_{21}^2$ and $\theta_{12}$ produce subleading effects on 
the atmospheric neutrino fluxes  for  $E >  1$ GeV. 
Moreover,  effect of these parameters in marginalization is very small  
due to their narrow allowed regions. 
Therefore we have taken fixed values of $\Delta m_{21}^2$ 
and $\theta_{12}$ in our analysis.

\begin{figure*}[htb]
\includegraphics[width=8.0cm,angle=0]{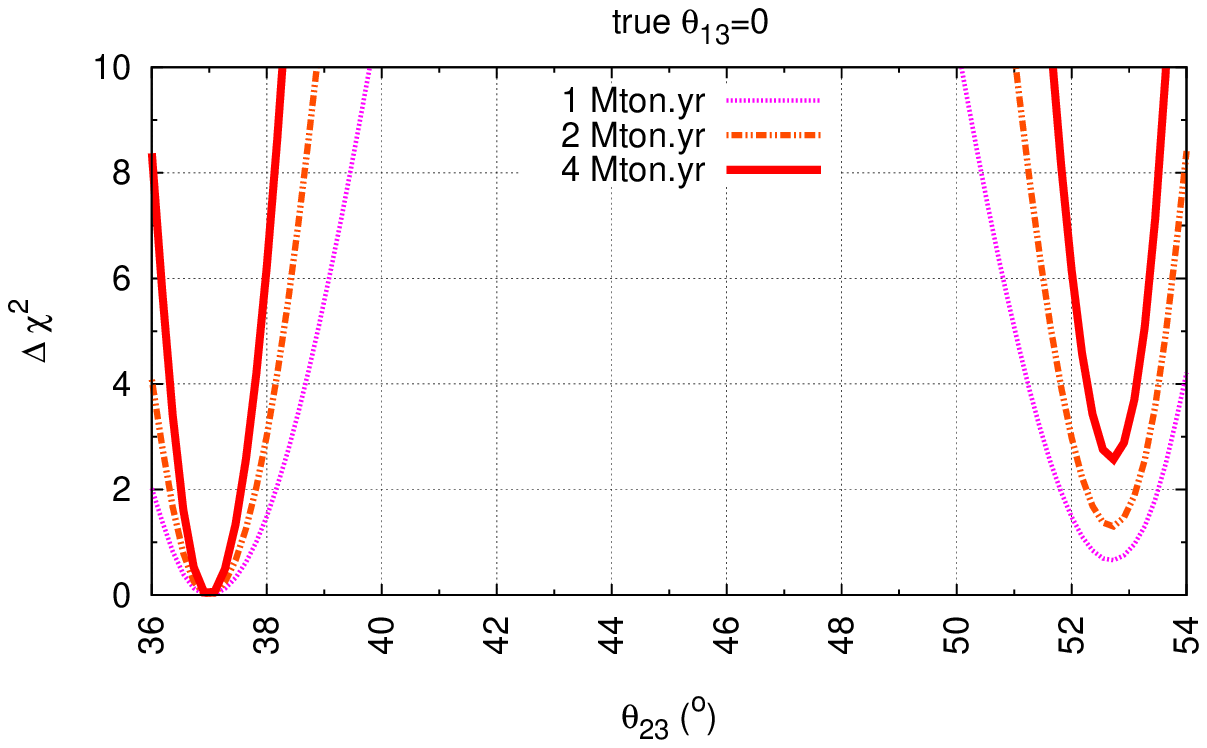}
\includegraphics[width=8.0cm,angle=0]{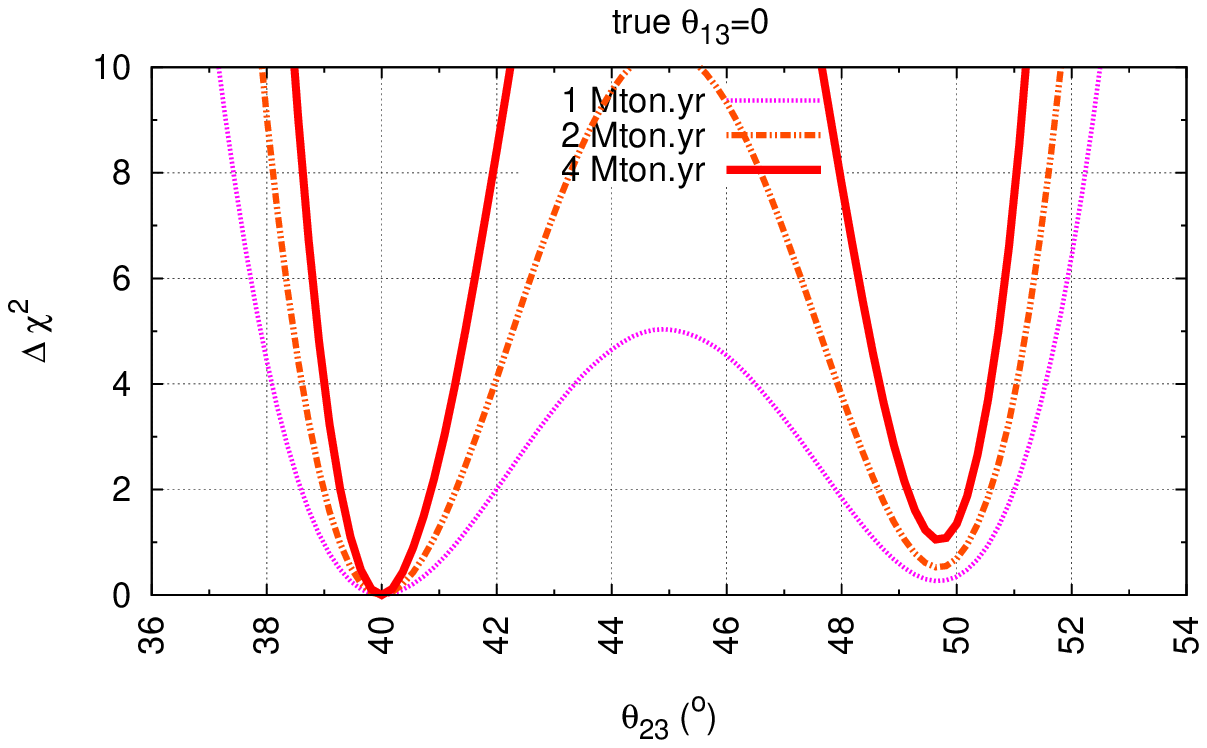}
\caption{\sf \small Dependence  of $\Delta \chi^2$  on fit value of 
$\theta_{23}$  for fixed input (true) values  $\theta_{23}=37^\circ$ (left) 
and $40^\circ$ (right).  
We used  $\theta_{13}=0$,  ${\mathcal E=}$ 1, 2, and 4 Mton$\cdot$yr, 
and the energy threshold  0.141 GeV. The $\chi^2$ is marginalized with respect to all 
the oscillation parameters except  $\theta_{23}$. 
The range of marginalization for $\theta_{13}$ is $0-10.5^\circ$.
}
\label{f:th23}
\end{figure*}

\begin{figure*}[htb]
\includegraphics[width=10.0cm,angle=0]{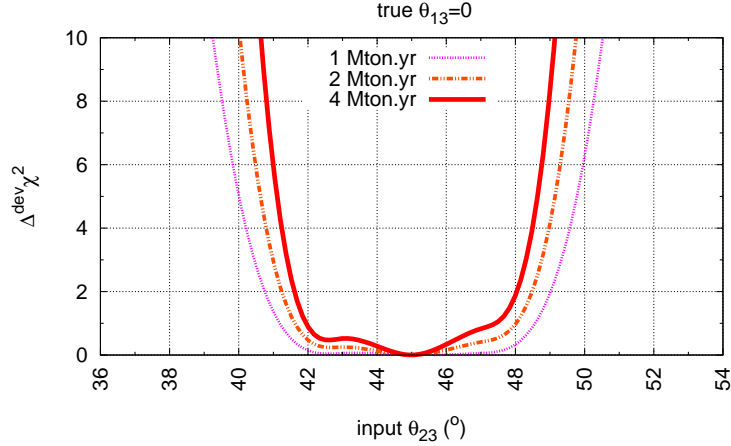}
\caption{\sf \small  
Dependence of the ICAL sensitivity  to the deviation from maximal mixing  
on the  input value of $\theta_{23}$
for  different exposure times: ${\mathcal E=}$ 1, 2, and 
4 Mton$\cdot$yr. We use the  threshold  0.141 GeV. 
Here $\Delta^{dev}\chi^2=\chi^2(45^\circ)-\chi^2(\theta_{23}^{true})$. 
$\chi^2$ is marginalized with respect to all the oscillation parameters except $\theta_{23}$.
The range of marginalization for $\theta_{13}$ is $0-10.5^\circ$.
}
\label{f:dmax}
\end{figure*}

\begin{figure*}[htb]
\includegraphics[width=10.0cm,angle=0]{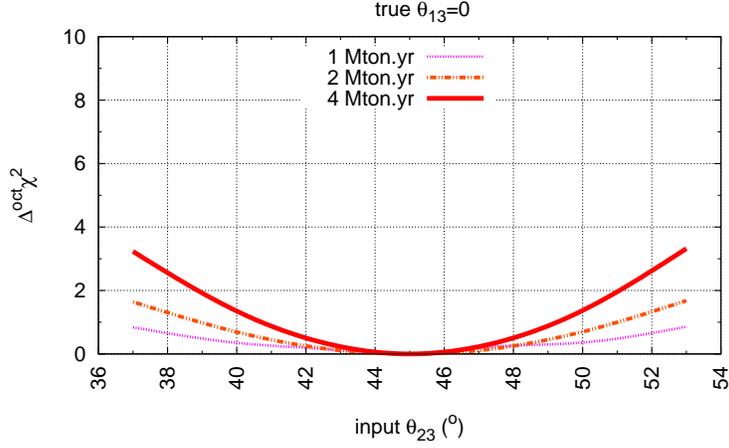}
\caption{\sf \small
Dependence of the ICAL sensitivity  to the $\theta_{23}$-octant on the 
input (true) value of $\theta_{23}$ for different values of ${\mathcal E}$, 
and  threshold 0.141 GeV. 
Here $\Delta^{oct}\chi^2 = \chi^2 (90^\circ - \theta_{23}) - \chi^2 (\theta_{23}).$
$\chi^2$ is marginalized with respect to all the oscillation parameters except $\theta_{23}$.
The range of marginalization for $\theta_{13}$ is $0-10.5^\circ$.
}
\label{f:oct}
\end{figure*}


\begin{figure*}[htb]
\includegraphics[width=10.0cm,angle=0]{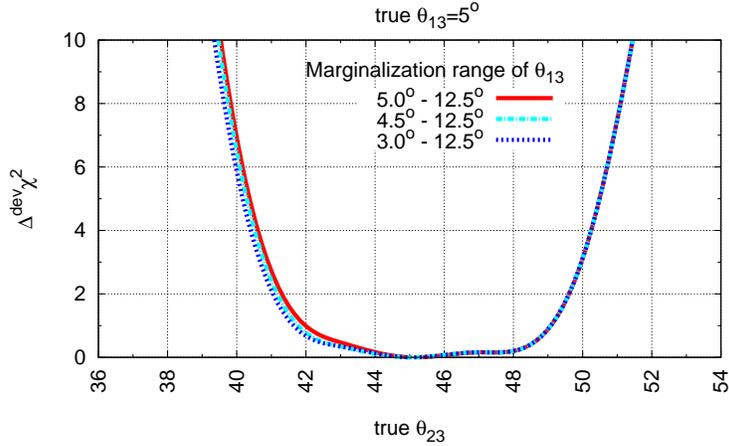}
\caption{\sf \small
Dependence  of the ICAL sensitivity  to the deviation from  maximal 2-3 mixing
on the input value of $\theta_{23}$ for different marginalization ranges
of $\theta_{13}$.  
We use 
${\mathcal E}=$ 1 Mton$\cdot$yr and threshold  0.806 GeV. 
Here $\Delta^{dev}\chi^2 \equiv \chi^2 (45^\circ) - \chi^2 (\theta_{23}^{true}).$
$\chi^2$ is marginalized with respect to
all oscillation parameters except $\theta_{23}$.
}
\label{f:dmax2}
\end{figure*}

\begin{figure*}[htb]
\includegraphics[width=10.0cm,angle=0]{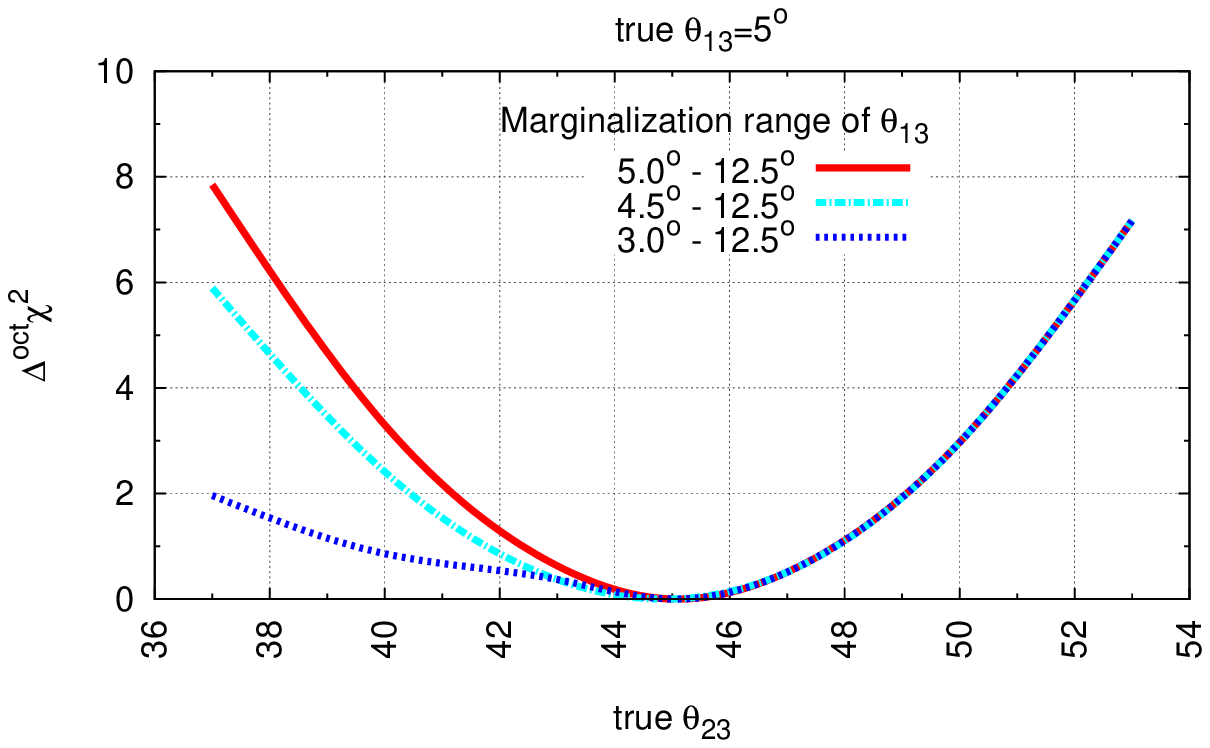}
\caption{\sf \small
Dependence of the ICAL sensitivity  to the $\theta_{23}$-octant on 
the input value of $\theta_{23}$  with ${\mathcal  E}=$ 1
Mton$\cdot$yr and threshold of 0.806 GeV. 
Here $\Delta^{oct}\chi^2 \equiv \chi^2 (90^\circ - \theta_{23}) - \chi^2 (\theta_{23}).$
$\chi^2$ is marginalized with respect to all
oscillation parameters except $\theta_{23}$.
}
\label{f:oct2}
\end{figure*}

\section{Sensitivities of ICAL}

In our computations we explored the neutrino energy range (0.141 -15) GeV, we used different 
energy thresholds and different exposures, ${\mathcal E}$, of 0.25, 1, 2, and 
4 Mton$\cdot$yr. 

\subsection{Determination of 2-3 mixing for $\theta_{13} = 0$}

The sensitivity of ICAL experiment to $\theta_{23}$ is shown in 
fig.~\ref{f:th23}.  We plot 
\be
\Delta 
\chi^2 \equiv \chi^2(\theta_{23}) - \chi^2(\theta^{true}_{23})
\label{eq:chi45}
\ee  
as function of the fit value 
for  fixed input values  $\theta_{23}^{true} = 37^\circ$,  
(left) and $40^\circ$ (right) with ${\mathcal E}=$ 1, 2, 
and 4 Mton$\cdot$yr. We have marginalized $\chi^2$ with respect 
to all the oscillation parameters  except $\theta_{23}$.
The figure shows high sensitivity to the deviation $\delta_{23}$:  
it would be possible to discriminate between a given $\theta_{23}$ and  maximal mixing 
at 99\% C.L., if $|\delta_{23}| > 5^{\circ}$. 
For instance, after 1 Mton$\cdot$yr the angle 
$\theta_{23}=37^{\circ}$ can be distinguished from $45^{\circ}$  at   $8\sigma$ level.

The figure shows also low sensitivity of the experiment to the octant.  
Indeed, $\Delta\chi^2$ is higher in the right minima  which correspond to 
the wrong octant.   After 1 Mton$\cdot$yr exposure the difference of
$\Delta \chi^2$ in the true and wrong octants is smaller than 1. 
The difference becomes more than 2 (90\% CL) only after 
4 Mton$\cdot$yrs. 
Identification of the octant becomes 
even more difficult for smaller $\delta_{23}$. 
If $\delta_{23} = 5^{\circ},$ we find $\Delta \chi^2 = 1.2$ for   
${\mathcal E} =  4$  Mton$\cdot$yr.
This result can be readily seen from our analytical consideration in sec.~II. 
The  probability $P_S$ averaged over the energy interval  $(0.14 - 15)$ GeV
equals $\langle P_S \rangle \sim 0.02$, so that for $\delta_{23} = 8^{\circ}$:  
$\Delta^{oct} F_\mu/ F^0_\mu \sim 4 \cdot 10^{-3}$, whereas 
$\Delta^{dev} F_\mu/ F^0_\mu \sim 0.04$ - an order of magnitude larger.  
As we mentioned before, this big difference of sensitivities  to the deviation and octant 
(degeneracy) is because the octant asymmetry is collected only at very 
low energies where $P_S$ is unsuppressed, whereas whole energy range contributes to 
the sensitivity to the deviation $\delta_{23}$. 

The sensitivity to  $\delta_{23}$ drops down substantially with 
decrease of   $\delta_{23}$. Reducing $\delta_{23}$ from $8^{\circ}$ to $5^{\circ}$ 
(compare the left and right panel of fig.~\ref{f:th23})
leads to decrease  of the flux difference of eq. (\ref{eq:deltadev})
by factor 2.5, and correspondingly, 
significance of discrimination from maximal mixing becomes 
$2 \sigma$ (for  ${\mathcal E} = 1$ Mton$\cdot$yr). 
Sensitivity to the octant at 
the $1 \sigma$ level appears only if ${\mathcal E} >  4$ Mton$\cdot$yr. 

In fig.~\ref{f:dmax} we show the marginalized $\Delta\chi^2$,
calculated at the maximal mixing (fit value) for different 
input (true) values of $\theta_{23}$: 
\be
\Delta^{dev} \chi^2 \equiv
\chi^2(45^{\circ}) - \chi^2(\theta_{23}^{true}).
\label{eq:chi45}
\ee
The picture is complementary 
to that in fig.~\ref{f:th23}, and there is an approximate 
symmetry  with respect to $\theta_{23}=45^\circ$.  
The reason of  sharp increase of $\Delta\chi^2$ at 42 and $48^{\circ}$ 
is related to weak dependence of the oscillation probability 
on $\delta_{23}=0$ around $\delta_{23}=0$  and 
to overall flux uncertainty.

In fig.~\ref{f:oct} we illustrate dependence of the sensitivity to the octant 
on  $\theta_{23}$. For different input (true) values of $\theta_{23}$ 
we plotted  
\be
\Delta^{oct}\chi^2 \equiv \chi^2 (90^\circ - \theta_{23}) - \chi^2 (\theta_{23}). 
\label{eq:chi90}
\ee
According to  fig.~\ref{f:oct}, it will  be possible to discriminate
the octant at 90\% CL if $\theta_{23}\lapp 38^\circ$ or  $\gapp 52^\circ$
after ${\mathcal E}=4$ Mton$\cdot$yr.
Notice that the curves are nearly symmetric with respect to 
$\theta_{23} = 45^{\circ}$.


Due to fast decrease of $P_S$ 
with increase of energy (see fig.~\ref{f:dhoct}) the sensitivity  
to the octant disappears for high values of the threshold.

\subsection{Determination of $\theta_{23}$ 
in the presence of non-zero 1-3 mixing.} 

Analysis of the oscillation data testifies for the non-zero 1-3
mixing, although significance of this result is not high and zero value 
of $\theta_{13}$ is not yet excluded. 
The global fit  at 1$\sigma$ C.L. gives
$ \theta_{13} = {7.3^\circ} \,^{+2.1^\circ}_{-3.2^\circ}$,
with 3$\sigma$ upper bound of $\theta_{13} < 13^\circ$, and
$\delta_\text{CP} \in [0,\, 360]$ \cite{GonzalezGarcia:2010er}.
Analysis of the solar and KAMLAND data by SNO collaboration leads to  
$\theta_{13} = {8.13^\circ} \,^{+3.53^\circ}_{-7.03^\circ}$ at 95\% CL \cite{Aharmim:2009gd}. 
New and forthcoming experiments
Double Chooz,  Daya Bay, RENO,  T2K, NO$\nu$A 
can confirm this result with higher confidence level or
put new stringent upper bound \cite{Mezzetto:2010zi}
which would  correspond approximately to a situation with 
zero 1-3 mixing considered in the previous section. 

By the time when ICAL will collect significant statistics 
the angle $\theta_{13}$ will be known with relatively good accuracy. 
To clarify an impact of this information on the determination of parameters 
of the 2-3 sector we have performed analysis for non-zero  $\theta_{13}$. 
For illustration purpose we use   $\theta_{13} = 5^{\circ}$ as the true value and 
different fit intervals with flat distribution 
(which could reflect errors in measurements of $\theta_{13}$). 

In fig.~\ref{f:dmax2} we show dependence of  $\Delta^{dev}\chi^2$ defined in eq.~(\ref{eq:chi45})
on the true (input) value of $\theta_{23}$  for the fit value   $\theta_{23} = 45^\circ$ 
and ${\mathcal E}=$ 1 Mton$\cdot$yr. 
Different curves correspond to different  marginalization intervals of $\theta_{13}$. 
Comparing these results with the results of fig.~\ref{f:dmax} we find that 
inclusion of the 1-3 mixing does not change significantly the sensitivity to 
$\delta_{23}$. The reason is that this sensitivity follows from 
the main mode of the $\nu_\mu-$oscillations; large 
probability for this mode  extends to 
higher energies and 1-3 mixing produces just additional distortion of 
the oscillatory pattern.  
However, inclusion of the 1-3 mixing makes the curve 
less symmetric with respect to $45^\circ $ which reflects 
an increase of sensitivity to the octant. 
Also sensitivity to the deviation weakly depends on the 
marginalization interval for $\theta_{13}$.   

Fig.~\ref{f:oct2} illustrates the sensitivity of ICAL to the 
octant in the presence of non-zero 1-3 mixing. 
We show dependence of $\Delta^{oct}\chi^2$ 
defined in eq.~(\ref{eq:chi90}) on the true value of  $\theta_{23}^{true}$ for the fit 
value $\theta_{23}^{fit} = 90^\circ-\theta_{23}$. Different curves correspond to different 
marginalization intervals for $\theta_{13}$.  
There are two important features of the fig.~\ref{f:oct2}. 
First, the sensitivity to the octant is substantially better  for non-zero 
value of $\theta_{13}$ than for vanishing 1-3 mixing (fig.~\ref{f:oct2})
as was also shown in previous publications \cite{GonzalezGarcia:2004cu,Hiraide:2006vh,Kajita:2006bt,Meloni:2008it}. 
This is related to the fact that the octant asymmetry 
of the flux is determined now by 
\be
\frac{\Delta^{oct} F}{F_\mu^0} \approx \sin 2 \delta_{23} \left( 1 - \frac{1}{r}\right) 
\langle P_A \rangle
\label{eq:ffA}
\ee 
 and  $\langle P_A \rangle \sim 0.1$  in the interval 
$E = (0.14 - 15)$ GeV; it  is much larger than  $\langle P_S \rangle$
being enhanced in the energy range  E = (3 - 10) GeV 
(in the resonance channel).   
Furthermore,  at  high energies $r \sim (3  - 4)$ and value of 
the coefficient in eq.~(\ref{eq:ffA}) becomes larger. 
As a result, for $\theta_{23} = 51^{\circ}$ the octant can be identified 
at $2\sigma$ level ($\Delta^{oct} \chi^2 = 4$) with  ${\mathcal E}=$ 1 Mton$\cdot$yr, 
as compared to $\Delta^{oct} \chi^2 = 0.3$ for $\theta_{13} = 0$. 

The second feature is significant asymmetry of the curves with 
respect to $\theta_{23} = 45^{\circ}$. The asymmetry is practically 
absent for fixed input value of  $\theta_{13}$  but it increases with 
broadening of the marginalization interval, and more importantly,  
with increase of the lower border of this interval. For   
$\theta_{23} > 45^{\circ}$ the curves  are practically not 
changed with change of the interval, whereas for $\theta_{23} < 45^{\circ}$ the sensitivity 
substantially decreases. For instance, taking $\theta_{23} = 40^{\circ}$ we obtain
$\Delta \chi^2 = 1$ for the interval
$\theta_{13}  = (3^{\circ} - 12.5^{\circ})$ instead of  $\Delta^{oct} \chi^2 = 3$ 
for fixed value  $\theta_{13}  = 5^{\circ}$. 

This asymmetry can be readily understood  from the analytic consideration 
of sec. II. 
Neglecting the effect of 1-2 mixing the $\nu_\mu-$flux can be presented  as 
\be
\frac{ F_\mu}{F_\mu^0} \approx  
K(\sin 2 \theta_{23}) - 
f (\theta_{23}) \left( 1 - \frac{1}{r}\right) P_A (\theta_{13}), 
\label{eq:flap}
\ee  
where  $K(\sin 2 \theta_{23}) $ is an even function of the 
 deviation (symmetric with respect to 
change of the octant), and  
\be
f (\theta_{23})  \equiv  \left(s_{23}^4 - \frac{s^2_{23}}{r} \right) 
\ee 
quickly increases with  $\theta_{23}$,  so that 
for $r =  3 - 4$  one has    $f (\theta_{23} = 40^{\circ}) \ll  
f (\theta_{23} > 50^{\circ})$. Therefore 
for  $\theta_{23} < 45^{\circ}$ the flux $F_\mu$
has much weaker  dependence on $\theta_{13}$ than for 
$\theta_{23} > 45^{\circ}$. In the process of marginalization 
over $\theta_{13}$ we compare the  true  value of the flux,  
$F_\mu^{true}$, with the  fit value, $F_\mu^{fit}$, 
and $\Delta^{oct}\chi^2$ is proportional  to their difference: 
\bea 
& & \frac{1}{F_\mu^0} \left| F_\mu^{true} -  F_\mu^{fit} \right| =  
 \left( 1 - \frac{1}{r}\right)  
\times
\nonumber \\     
&&\left| f (\theta_{23})  \langle P_A (\theta_{13}^{true}) \rangle  
- f (90^{\circ} - \theta_{23}) \langle P_A (\theta_{13}) \rangle  \right| . 
\label{eq:fchi}  
\eea
If the fit value of  1-3 mixing 
is fixed: $\theta_{13}^{fit} = \theta_{13}^{true}$,
the curve is approximately symmetric with respect to 
change $\theta_{23} \leftrightarrow (90^{\circ} -  \theta_{23})$. Indeed, 
in this case we have from eq.~(\ref{eq:fchi})
\bea 
&& \frac{1}{F_\mu^0} \left| F_\mu^{true} -   F_\mu^{fit} \right|  =
\left( 1 - \frac{1}{r}\right)
\nonumber\\ && 
\times
\langle P_A (\theta_{13}^{true}) \rangle
\left| f (\theta_{23}) - f (90^{\circ} - \theta_{23}) \right| . 
\label{eq:symm}  
\eea
The situation is different if  $\theta_{13}$  varies in  
certain interval  $\theta_{13} = [\theta_{13}^{min} - \theta_{13}^{max}]$,   
and we perform marginalization over $\theta_{13}$ in this interval. 
Marginalization minimizes the difference of fluxes (eq. (\ref{eq:fchi})) over 
$\theta_{13}$ for a given value of  $\theta_{23}$. 
If  $\theta_{23} < 45^{\circ}$, then  
$f(\theta_{23})  \ll f(90^{\circ} -  \theta_{23})$. 
In this case the difference of fluxes is minimal if   
$\theta_{13} \sim \theta_{13}^{min}$. Indeed, 
since $\langle P_A \rangle$ decreases with $\theta_{13}$,  
a small value of  $\langle P_A \rangle$  partially compensates large value of  
$f(90^{\circ} -  \theta_{23})$ in the second 
term on the right hand side of eq.~(\ref{eq:fchi}). 
Furthermore, the smaller $\theta_{13}^{min}$ the stronger compensation,  
and therefore the smaller $\Delta\chi^2$ can be obtained. 
If  $\theta_{23} > 45^{\circ}$, then  
$f(\theta_{23})  \gg  f(90^{\circ} -  \theta_{23})$.  
Now to compensate  the first term in eq.~(\ref{eq:fchi}) one should take  
$\langle P_A(\theta_{13}^{fit})  \rangle \gg 
\langle P_A(\theta_{13}^{}) \rangle$. This is, however, not possible 
for the considered values of  $\theta_{13}$.  
Thus, in the case of unprecise determination of 
$\theta_{13}$  sensitivity to the octant is higher 
for $\theta_{23} > 45^{\circ}$.

\begin{figure*}[htb]
\includegraphics[width=8.0cm,angle=0]{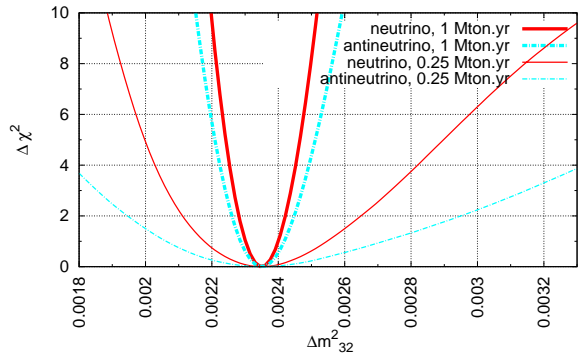}
\caption{\sf \small Dependence of $\Delta \chi^2$ on the fit value 
of  $\Delta m_{32}^2$ for the true value $\Delta m_{32}^2 = 2.35 \times 10^{-3}$ eV$^2$
in the neutrino and antineutrino channels. We use   
${\mathcal E}=$ 0.25 and 1 Mton$\cdot$yr and the  threshold 0.8 GeV. 
The marginalization is done over all the oscillation
parameters except  $\Delta m_{32}^2$.
The  $\chi^2$ values 1, 4, 9  correspond to 1$\sigma$ (68.3\%), 2$\sigma$ (95.4\%), 
and 3$\sigma$ (99.73\%), respectively.
}
\label{f:dmsq}
\end{figure*}

\begin{figure*}[htb]
\includegraphics[width=8.0cm,angle=0]{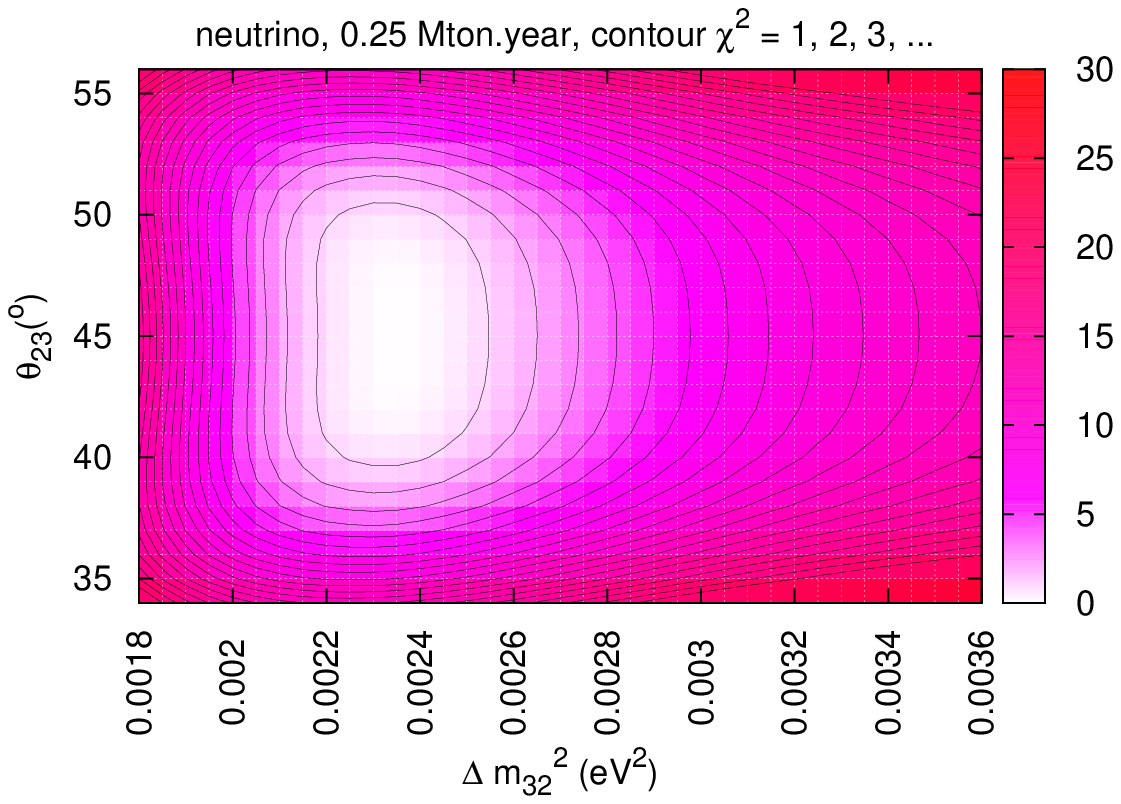}
\includegraphics[width=8.0cm,angle=0]{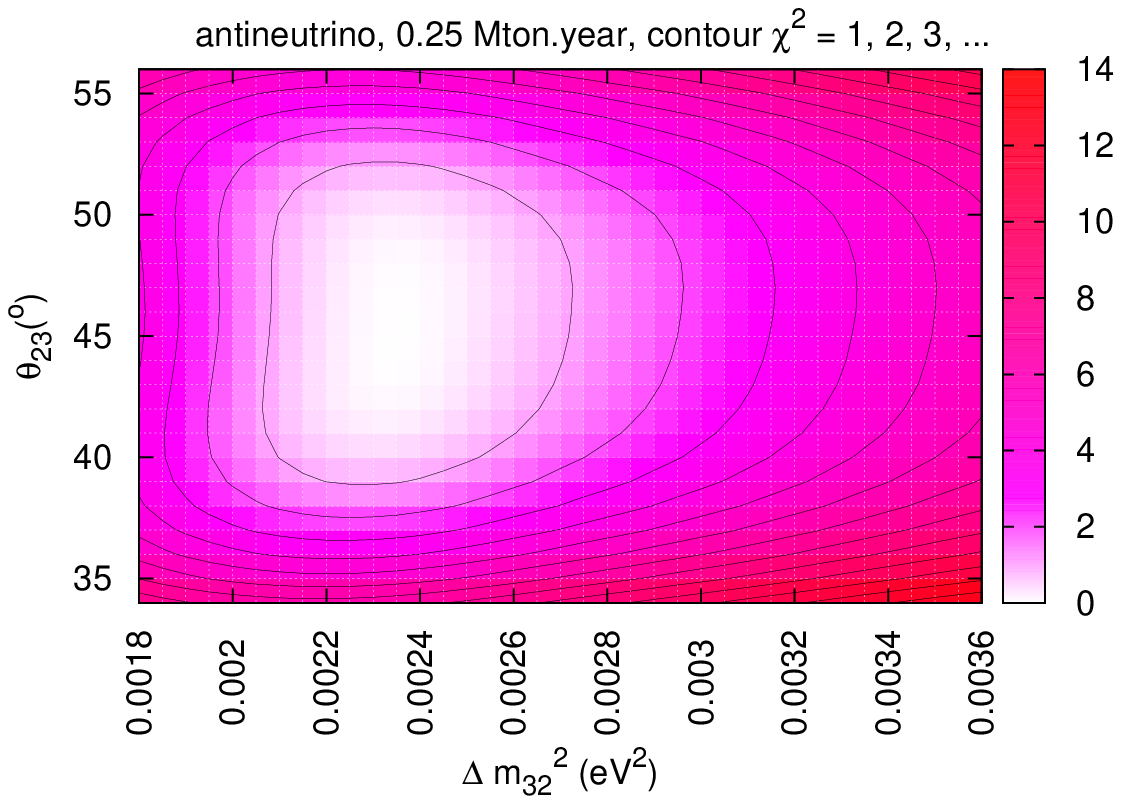}
\caption{\sf \small The iso-$\chi^2$  contours 
in the $\Delta m^2_{32} - \theta_{23}$ plane   for neutrinos (left)
and antineutrinos (right) obtained with  ${\mathcal E}=$ 0.25 Mton$\cdot$yr. 
The inner contour corresponds to  $\chi^2=1$ and others  with increment 1.  
We use the input value $\Delta m^2_{32} =-2.35 \times 10 ^{-3}$eV$^2$, $\theta_{23}=45^\circ$,  
$\theta_{13}=5^\circ$ and the threshold 0.8 GeV.
Here  $\Delta\chi^2$ values 2.3, 4.6, 9.2  correspond to 1$\sigma$ (68.3\%), 2$\sigma$ (90\%), 
and 3$\sigma$ (99\%), respectively. Intensity of color reflects value of $\Delta\chi^2$ 
(see column on the right hand sides of the panels for identification). 
}
\label{f:dmsqthy5}
\end{figure*}

\begin{figure*}[htb]
\includegraphics[width=8.0cm,angle=0]{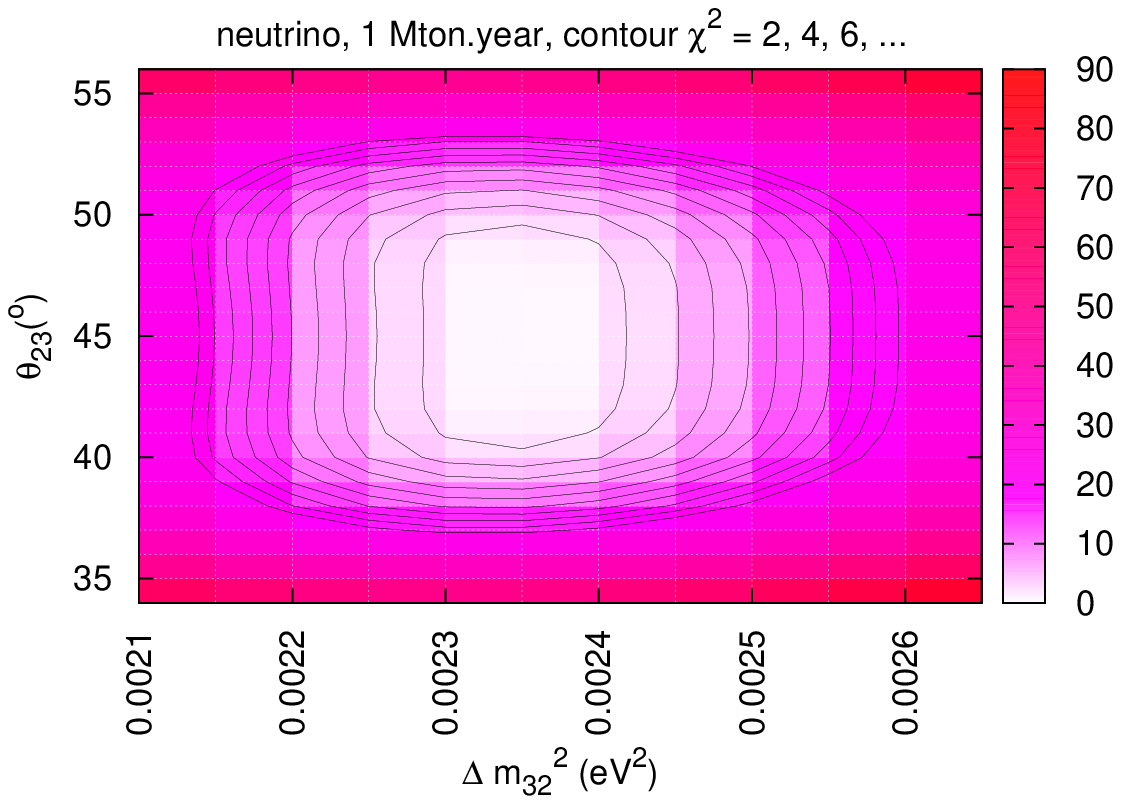}
\includegraphics[width=8.0cm,angle=0]{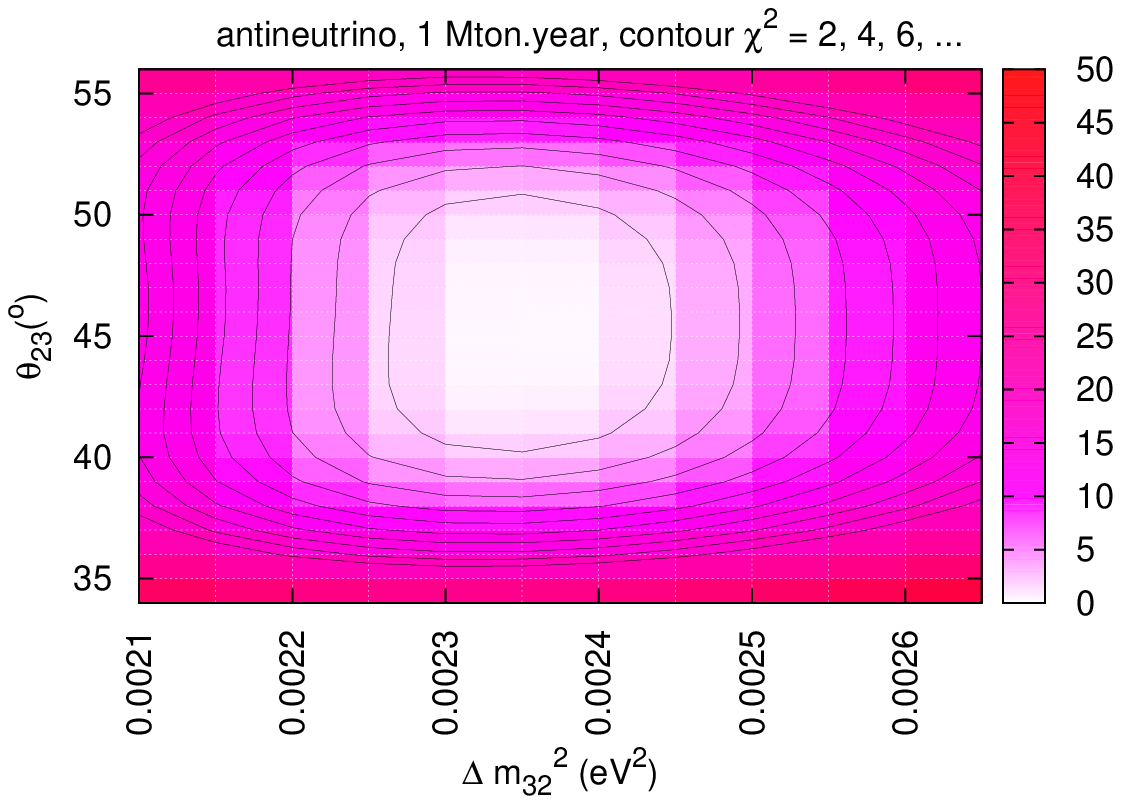}
\caption{\sf \small The same as in  fig. \ref{f:dmsqthy5}  but 
for  ${\mathcal E}=$ 1 Mton$\cdot$yr. 
The inner contour corresponds to $\chi^2=2$  and 
for other contours the increment   $\Delta \chi^2 = 2$.
}
\label{f:dmsqthy20}
\end{figure*}

\subsection{Determination of the 2-3 mass split. CPT test} 

Important advantage of a  magnetized calorimeter is that it allows 
one to measure the neutrino mass differences and mixing angles in the neutrino 
 and antineutrino channel separately. A difference of results
can be related to some effective or fundamental 
violation of the CPT symmetry.  

In fig.~\ref{f:dmsq} we show dependence of $\Delta\chi^2$ on the fit value  
of  $\Delta m^2_{32}$ for the true value 
$\Delta m_{32}^2 = 2.35 \cdot 10^{-3}$ eV$^2$ 
in the neutrino and antineutrino channels.  We take $\theta_{13} = 5^\circ$ and 
$\theta_{23} = 45^{\circ}$. 
According to this figure  with ${\mathcal E}= 0.25$  Mton$\cdot$yr  the value 
$\Delta m_{32}^2 = 3.3 \cdot 10^{-3}$ eV$^2$ can be discriminated from the true value at 
about 2$\sigma$ level.

The accuracy of measurement of $\Delta m_{32}^2$ is better in  the  
$\nu-$channel. For  ${\mathcal E}= 0.25$ Mton$\cdot$yr the error in $\nu$-channel 
is about two times smaller than that in $\bar\nu$ channel. The difference 
of accuracies decreases with increase of exposure and e.g. 
for  1 Mton$\cdot$yr it becomes about 25\%.
The curves $\Delta\chi^2$ are asymmetric with respect to ${\Delta m_{32}^2}^{true}$,
which is related to the dependence of oscillation probability on $\Delta  m_{32}^2$.

The $1 \sigma$ error for $\Delta m^2_{23}$ could be $0.15 \cdot 10^{-3}$ eV$^2$ 
and $0.04 \cdot 10^{-3}$ eV$^2$ after 
0.25 Mton$\cdot$yr and 1 Mton$\cdot$yr exposures correspondingly. 
With  ${\mathcal E} = 1$ Mton$\cdot$yr the error  
$0.15 \cdot 10^{-3}$ eV$^2$ can be achieved at $3\sigma$ 
level. With ${\mathcal E}=  1$ Mton$\cdot$yr  one can  obtain 
an accuracy $0.08 \cdot 10^{-3}$ eV$^2$ ($90 \%$ C.L.)   
which is better than the present MINOS accuracy.

In figs.~\ref{f:dmsqthy5} and \ref{f:dmsqthy20} we show $\Delta\chi^2$ as function of  
$\Delta m_{32}^2$ and  $\theta_{23}$ in the neutrino and antineutrino channels 
for  ${\mathcal E}=$ 0.25 and 1 Mton$\cdot$yr.
As true values we take  $\theta_{23} = 45^{\circ}$ 
and $\Delta m_{32}^2 = 2.35 \cdot 10^{-3}$ eV$^2$. 
According to  fig.~\ref{f:dmsqthy20} the present MINOS result for $\bar\nu$
($|\Delta m_{31}^2|=3.36\cdot  10^{-3}~{\rm eV}^2,$
 $\theta_{23} = 34^\circ$) 
can be excluded by ICAL at about 
$6\sigma$ level with  ${\mathcal E}=$  1 Mton$\cdot$yr.

\section{Further improvements of sensitivities}

There are several directions in which 
sensitivity of ICAL can be further improved. 

\subsection{Adding information about hadrons} 

\begin{figure*}[htb]
\includegraphics[width=8.0cm,angle=0]{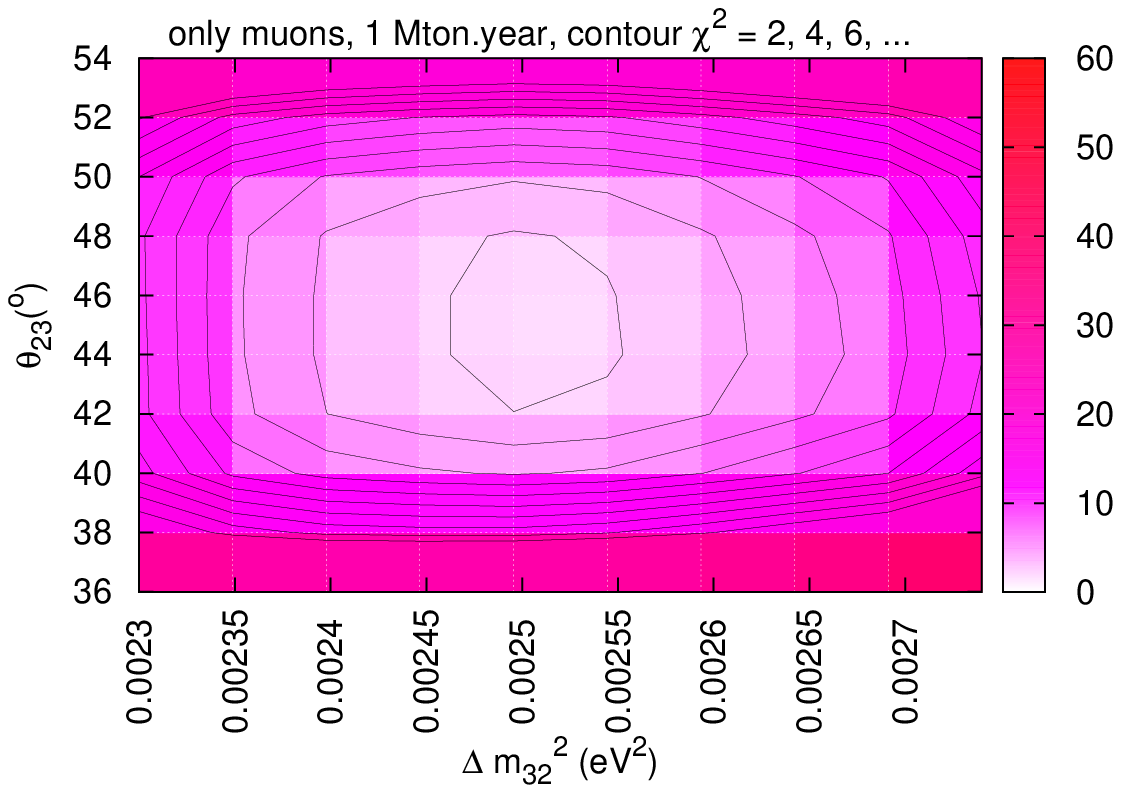}
\includegraphics[width=8.0cm,angle=0]{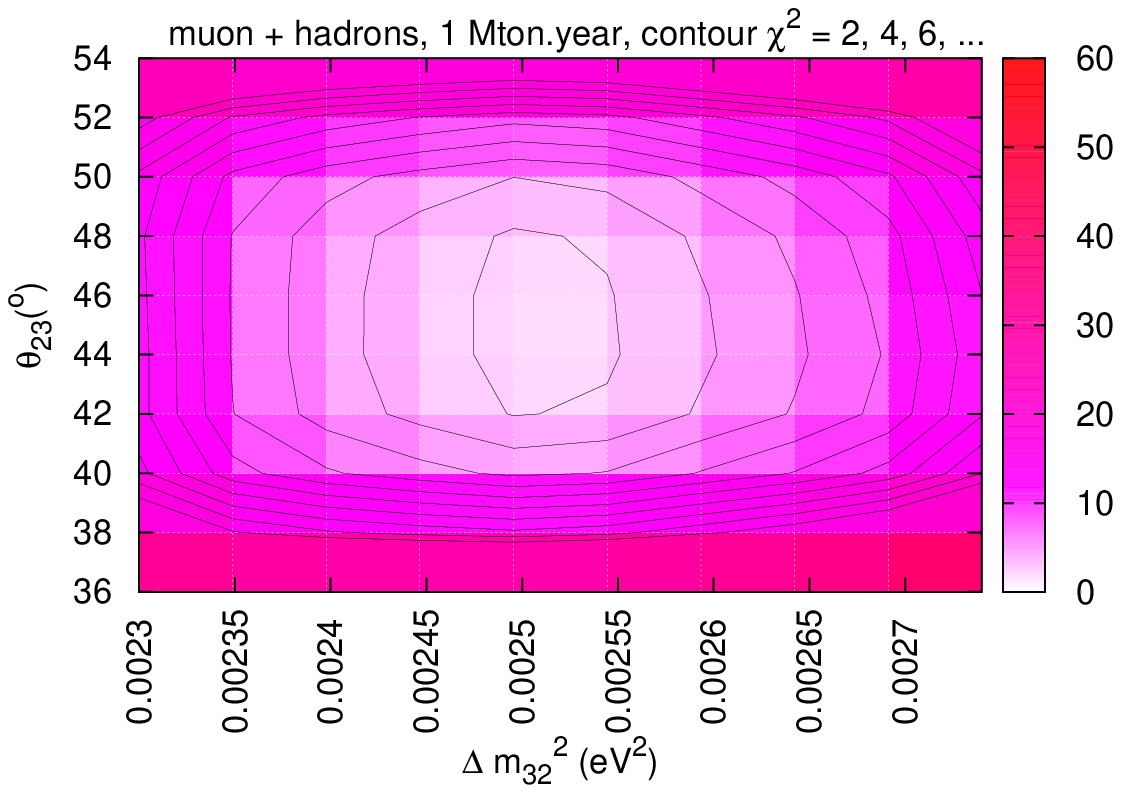}
\caption{\sf \small Effect of inclusion in the analysis the information on hadrons. 
The iso-$\chi^2$  contours (from inner side $\chi^2=2$ with increment 2) 
in the $\Delta m^2_{32} - \theta_{23}$ plane  without hadrons (left) and with hadrons (right). 
We take 1 Mton$\cdot$yr,  the threshold 0.8 GeV for both muons and hadrons.
and sum up the signals from neutrinos and antineutrinos.  
}
\label{f:dmsqthad}
\end{figure*}

Measurements  of the hadron  
energy in ICAL in addition to the muon energy is expected to improve 
reconstruction of the neutrino energy  for $E\gapp 2$ GeV. 
However, the total hadron energy in an event is carried out by multiple 
low energy hadrons. The average energy per hadron per event is 
$\lapp 1$GeV and the average number of hadrons per event is $\gapp 2$ 
At $E \lapp 1$GeV  the energy resolution is very poor,  
roughly $80\%$, and the number of hits 
(number of active detector layers in which signal is detected) 
increases only logarithmically with $E$.

The resolution of hadron energy (for all pions and kaons) 
at ICAL has been obtained from GEANT4 simulation and 
parametrized as 
\begin{equation}
\sigma_{had}/E_{had}=a/\sqrt{E_{had}}+b,
\label{e:sig_had}
\end{equation}
where, $a \approx 0.60$ and $b\approx0.1$
for the thickness of iron layer  5.6 cm and after averaging over all directions.
 
For each hadron in an event we  find the reconstructed hadron energy ($E_{had}^{rec}$)
by a random number method using the value of $\sigma_{had}$ from eq. \ref{e:sig_had}. 
Then we find the final resolution as a function of $[E_\nu - (E_\mu + E_{had}^{recS})]/E_\nu$;
where, $E_{had}^{recS}$ is the sum of the energies of all reconstructed hadrons in 
an event. We use  the atmospheric neutrino events without oscillations 
for an exposure of $50 \times 1000$ kTon$\cdot$yr generated by Nuance to reconstruct   
the resolution functions for each energy and zenith angle bins.

In fig. \ref{f:dmsqthad} we show the iso-$\chi^2$ contours with and without 
inclusion of information about hadrons.
We find that improvement of the sensitivity to $\theta_{23}$,  
and therefore $\delta_{23}$, is marginal. 
Also there is a very small change  in the sensitivity to $\Delta m_{32}^2$: 
$\Delta(\Delta m_{32}^2) = 0.02 \cdot 10^{-3}$ eV$^2$.

\subsection{Cross-sections and fluxes}

\begin{figure*}[htb]
\includegraphics[width=10.0cm,angle=0]{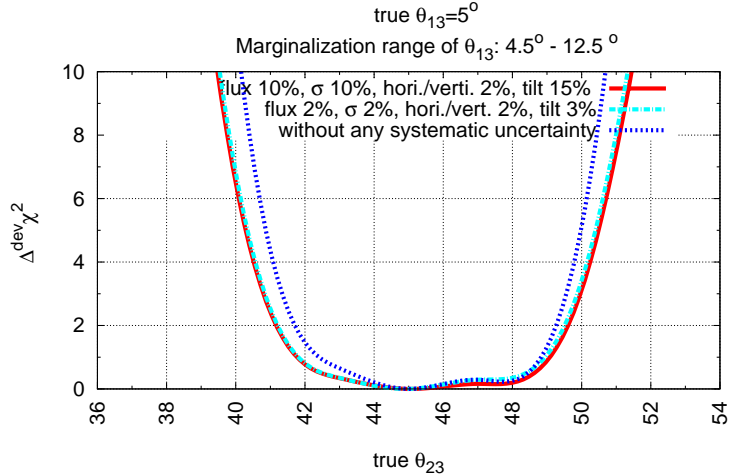}
\caption{\sf \small
The same as in  fig.~\ref{f:dmax2},  but with 
reduced  systematic uncertainties.
}
\label{f:dmax_uncer}
\end{figure*}

\begin{figure*}[htb]
\includegraphics[width=10.0cm,angle=0]{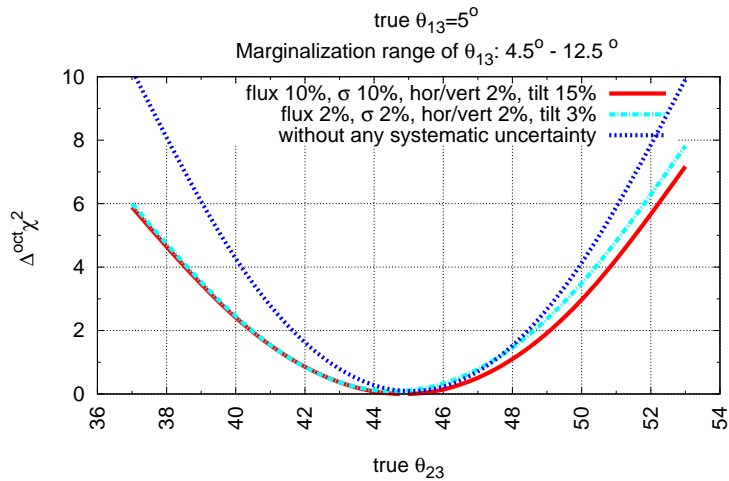}
\caption{\sf \small
The same as in  fig.~\ref{f:oct2},  but with reduced systematic
uncertainties.
}
\label{f:oct_uncer}
\end{figure*}

Figs.~\ref{f:dmax_uncer} and \ref{f:oct_uncer} illustrate improvements 
of the sensitivities with
reduction of different systematic uncertainties. Decrease of uncertainties of the  
overall flux normalization, the ratio of horizontal/vertical flux, the neutrino  cross-section, 
and the tilt (below 1 GeV)  from 10\%, 10\%, 2\%, and 15\%  to 
 2\%, 2\%, 2\%, and 3\%, respectively, do not lead  to significant improvement of the sensitivities.
The reason is that  the up-going neutrinos  oscillate and down-going neutrinos
remain practically unchanged. In $\chi^2$ analysis these down-going neutrinos
allow to reduce the effect of systematic uncertainties in fluxes. Significant improvement
occurs only when all systematic uncertainties are zero.

\subsection{Adding priors to $\chi^2$}

In  $\chi^2$ analysis in sec. 4  
for simplicity  we used flat distributions 
of uncertainties of the oscillation parameters
in  marginalization. 
The prior contribution of the parameters to $\chi^2$. 
should  improve the sensitivity. Proper procedure would require the use of the best-fit values 
and variations of parameters (especially $\theta_{13}$)
which will be possible after results of forthcoming experiments will be known.

\begin{figure*}[htb]
\includegraphics[width=10.0cm,angle=0]{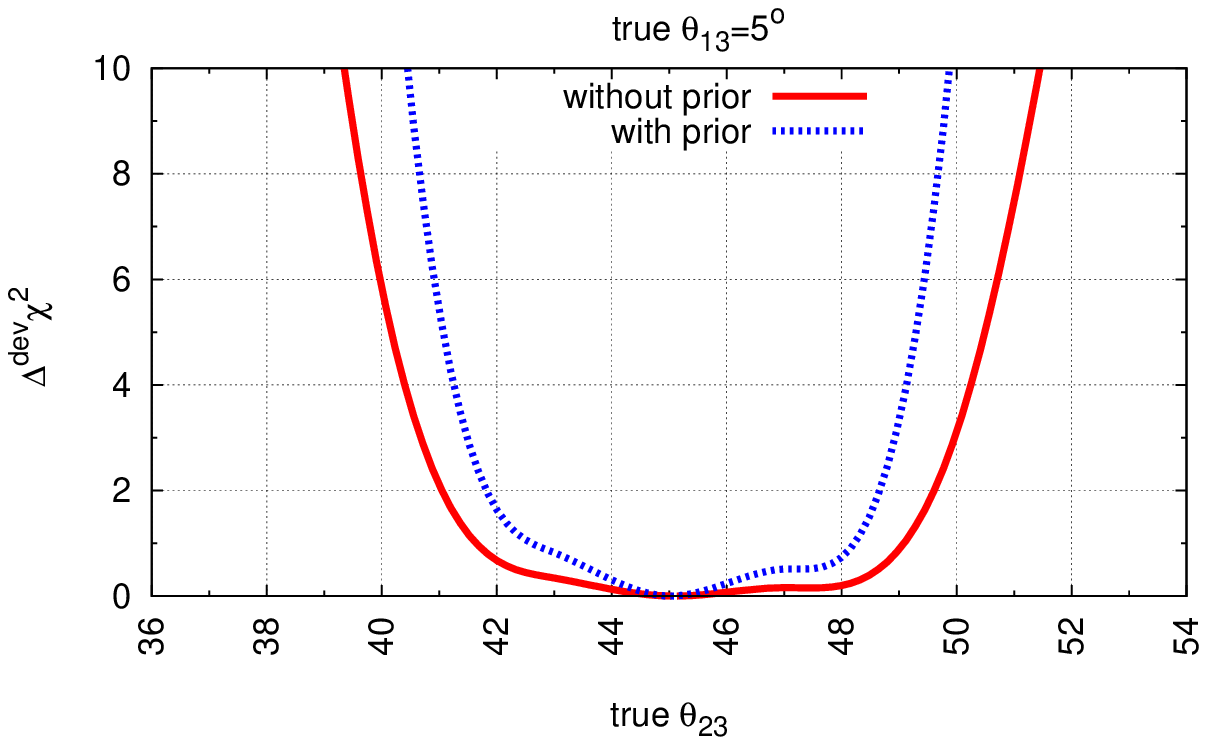}
\caption{\sf \small
The same as in  fig.~\ref{f:dmax2},  but with the contributions  from priors 
for the oscillation parameters added to $\chi^2$ (see text). 
The marginalization range for $\theta_{13}$ is $3^\circ - 12.5^\circ 
(0^\circ - 12.5^\circ)$ in absence  (presence) of prior contribution.
}
\label{f:dmax_prior}
\end{figure*}

\begin{figure*}[htb]
\includegraphics[width=10.0cm,angle=0]{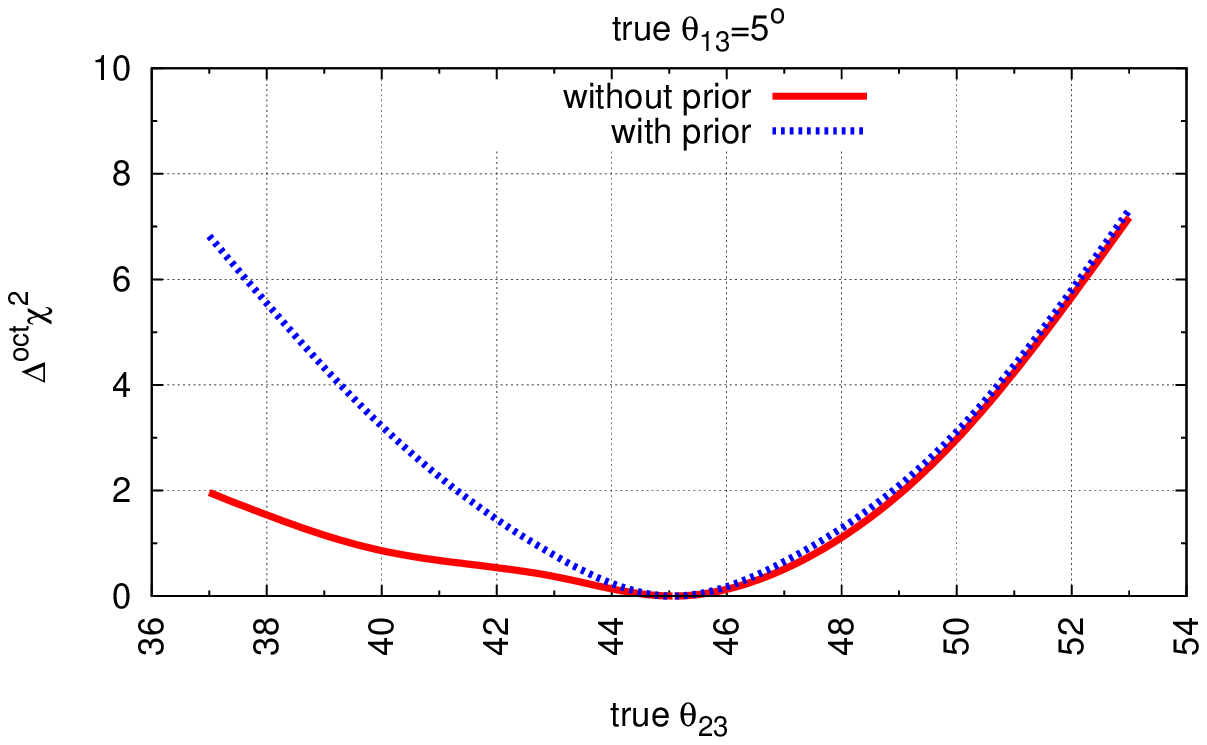}
\caption{\sf \small
The same as in  fig.~\ref{f:oct2},   but  with the contribution 
from priors for the  oscillation parameters added to the $\chi^2$. 
The marginalization range for $\theta_{13}$ is $3^\circ - 12.5^\circ 
(0^\circ - 12.5^\circ)$  in absence (presence) of prior contribution.}
\label{f:oct_prior}
\end{figure*}

In figs.~\ref{f:dmax_prior} and \ref{f:oct_prior} we show improvements 
of the sensitivities to the $\theta_{23}$ deviation 
and the octant due to inclusion of  the  prior contribution.  
We have assumed the Gaussian distribution of the uncertainties around the best fit with width
$\sigma(\sin^22\theta_{13}) = 0.01$,  $\sigma (\sin^22\theta_{23}) = 0.015$ 
and $\sigma (\Delta  m_{32}^2)/ \Delta m_{32}^2=0.015$ following \cite{Huber:2009cw,Huber:2004ug}.  
The $2\sigma$ errors in measurements of $\Delta  m_{32}^2$ are 
$\pm 0.09$  with flat uncertainties of oscillation parameters,   
$\pm 0.07$ with prior information from present global-fit, and 
$\pm0.018$  with prior information from possible T2K result.

The asymmetry in the sensitivity to the octant (which was due to the uncertainty of 
$\theta_{13}$ in absence of prior contribution)  
now disappears, and the result does not depend on marginalization range. 

Of course, larger values of $\theta_{13}$ can substantially enhance the sensitivity to 
the octant since $P_A \sim \sin^2\theta_{13}$ at low energies.


\begin{table}[t]
\caption{Results of determination of $\theta_{23}$ 
}
\label{tab:th23}
\centering
\begin{tabular}{lrr}
\hline
\hline
$\theta_{23}$ & CL & Source \\
\hline
\hline
\vspace*{.3cm}
${42.9^\circ}^{+4.1^\circ}_{-2.8^\circ}$  & 1$\sigma$& global-fit\cite{GonzalezGarcia:2010er}  \\
\vspace*{.2cm}
$35.7^\circ - 54^\circ$  & 3$\sigma$  & global-fit\cite{GonzalezGarcia:2010er}\\
\vspace*{.2cm}
${45^\circ}^{+10^\circ}_{-7.8^\circ}$ & 99\% &SK \cite{Hosaka:2006zd}\\
\vspace*{.2cm}
$45^\circ\pm9^\circ$ & 90\% & MINOS ($\nu$) \cite{Hosaka:2006zd}\\
\vspace*{.2cm}
${34^\circ}^{+6^\circ}_{-4^\circ}$ or ${56^\circ}^{+4^\circ}_{-6^\circ}$ & 90\% & MINOS ($\bar\nu$) \cite{minos}\\ 
\vspace*{.2cm}
$39^\circ-51^\circ$& 2$\sigma$ & T2K \cite{Huber:2004dv}\\
\vspace*{.2cm}
$36^\circ - 54^\circ$ & 2$\sigma$ & NO$\nu$A \cite{Huber:2004dv}\\
\vspace*{.2cm}
$40^\circ-50^\circ$ & 2$\sigma$ & INO (1 Mton$\cdot$yr)\\
\hline
\hline
\end{tabular}
\end{table}

\begin{table}[t]
\caption{Results of determination of $\Delta m^2_{31}$ 
}
\label{tab:dmsq}
\centering
\begin{tabular}{lrr}
\hline
\hline
$\Delta m_{32}^2 (10^{-3}$eV$^2$) & CL & Source \\
\hline
\hline
\vspace*{.3cm}
$-2.36\pm 0.07(\pm 0.36)$ & 1 (3)$\sigma$ & global-fit \cite{GonzalezGarcia:2010er}\\
\vspace*{.2cm}
$+2.47\pm0.12(\pm0.37)$ &  1 (3)$\sigma$ & global-fit \cite{GonzalezGarcia:2010er}\\
\vspace*{.2cm}
$2.5^{+0.52}_{-0.60}$ & 99\% & SK 3$\nu$ \cite{Hosaka:2006zd}\\
\vspace*{.2cm}
$2.35^{+0.11}_{-0.08}$ & 90\%  & MINOS $\nu$ \cite{minos}\\
\vspace*{.2cm}
$3.36^{+0.45}_{-0.40}$ & 90\%  & MINOS $\bar\nu$ \cite{minos}\\
\vspace*{.2cm}
$2.5\pm 0.04 $ & 2$\sigma$ &T2K  \cite{Huber:2004dv}\\
\vspace*{.2cm}
$ 2.5_{-0.04}^{+0.07}$ & 2$\sigma$ &  NO$\nu$A \cite{Huber:2004dv}  \\
\vspace*{.2cm}
$2.5\pm0.07$ & 2$\sigma$ & INO (1 Mton$\cdot$yr)\\
\hline
\hline
\end{tabular}
\end{table}

\section{Conclusion}

We have studied analytically 
the dependence of the $\theta_{23}-$deviation effect and  octant asymmetry 
of $\nu_\mu$ and $\bar\nu_\mu$ fluxes on 
the neutrino parameters $\theta_{23}$ and $\theta_{13}$. 
We explored  numerically a sensitivities of a magnetized calorimeter 
to the $\theta_{23}$-deviation, to the octant and  to $\Delta m_{32}^2$. 

We show that for $\theta_{13} = 0$ the sensitivity of ICAL  to 
the  octant is  low even for maximally allowed values of 
the deviation of the 2-3 mixing from maximal.  
This is related to the fact that the octant asymmetry is 
proportional to the ``solar'' probability   $P_S$ which is  large: $O(1)$ 
at $E \sim 0.1$ GeV but quickly, as $\propto  E^{-2}$, decreases 
with energy. The situation can be improved by lowering the threshold, 
increasing exposure and reducing  systematic errors (especially 
in spectral index).  
We find that  $sign (\delta_{23})$ (octant) can be established at $90\%$ C.L. 
if $|\delta_{23}| = 7^\circ$,  ${\mathcal E}=$  4 Mton$\cdot$yr and $E_{th} = 0.141$ GeV.

ICAL has good sensitivity to the $\theta_{23}$-deviation from maximal 2-3 
mixing: the effect is proportional to the probability of the main channel  
of oscillations,  $\nu_\mu - \nu_\tau$, 
which is unsuppressed in whole considered neutrino energy range. 
As a result, dependence of the sensitivity on 
the energy threshold is weak and it does not change substantially 
when the effect of 1-3 mixing is included. 
We find that with  the 1 Mton$\cdot$yr exposure the $3\sigma$ accuracy of
determination of the deviation will be $|\delta_{23}| \approx 6^{\circ}$, 
which is better than the present global fit  result and slightly better than expected 
sensitivity of T2K ($\approx 9^{\circ}$).

The oscillations driven by non-zero 1-3 mixing  substantially improve 
the sensitivity  to the octant.  One can determine
the octant  for $\delta_{23} =5^\circ$ and $\theta_{13}=5^\circ$ at 90\% C.L. 
with  1 Mton$\cdot$yr exposure.  
We  find that this sensitivity depends crucially on the uncertainty range of  
$\theta_{13}$. For a given nonzero $\theta_{13}$,  the
sensitivity to octant discrimination is symmetric in $\theta_{23}$ 
with respect to $\theta_{23}=45^\circ.$  However, the asymmetry arises 
(smaller sensitivity for $\theta_{23} < 45^\circ$) if  value
of $\theta_{13}$ can vary in large range. 
The symmetry is restored if prior for the 1-3 mixing is added.

The accuracy of measurements of $\Delta m_{23}^2$ by ICAL,   
$\Delta (\Delta m_{23}^2) =  0.15 \cdot 10^{-3}$
eV$^2$ ($3\sigma$, 1 Mton$\cdot$yr exposure), is two times 
better than  the accuracy of the present global fit and it 
is worthier than the expected sensitivity of T2K. 

ICAL can measure the difference of  $\Delta m_{32}^2$ in $\nu$ and $\bar\nu$ channels 
(the CPT test) with accuracy  $0.8\times 10^{-4}$ eV$^2$ at 3$\sigma$ confidence level with 1 
Mton$\cdot$yr exposure and the present MINOS result can be excluded 
at  $>5\sigma$ confidence level.

We find that inclusion of information about hadrons from neutrino interactions 
does not change the sensitivity to the oscillation parameters substantially. 
Also improvements of the sensitivity due to better determinations  
of the cross-section and neutrino flux are  
rather modest. However, the sensitivity improves substantially with adding  priors, especially 
for the 1 - 3 mixing, and of course,  with increase of the absolute value of 1-3 mixing. \\

{\it Acknowledgments:}
The use of general cluster facility of Harish-Chandra Research Institute  
for a part of this work is gratefully acknowledged.
A. S. wants to acknowledge Moon Moon Devi and Amol Dighe 
for providing the hadron resolution function.


\begin{thebibliography}{99}

\bibitem{Mohapatra:2006gs}
  R.~N.~Mohapatra and A.~Y.~Smirnov,
  Ann.\ Rev.\ Nucl.\ Part.\ Sci.\  {\bf 56}, 569 (2006)
  [arXiv:hep-ph/0603118]; and the reference there in.


\bibitem{GonzalezGarcia:2010er}
  M.~C.~Gonzalez-Garcia, M.~Maltoni and J.~Salvado,
  JHEP {\bf 1004}, 056 (2010)
  [arXiv:1001.4524 [hep-ph]].

\bibitem{Fogli:2008jx}
  G.~L.~Fogli, E.~Lisi, A.~Marrone, A.~Palazzo and A.~M.~Rotunno,
  Phys.\ Rev.\ Lett.\  {\bf 101}, 141801 (2008)
  [arXiv:0806.2649 [hep-ph]].


\bibitem{Hosaka:2006zd}
  J.~Hosaka {\it et al.}  [Super-Kamiokande Collaboration],
  Phys.\ Rev.\  D {\bf 74}, 032002 (2006)
  [arXiv:hep-ex/0604011].

\bibitem{minos} 
  P.~Adamson {\it et al.} [ The MINOS Collaboration ],
  [arXiv:1103.0340 [hep-ex]].


\bibitem{Kopp:2010qt}
  J.~Kopp, P.~A.~N.~Machado and S.~J.~Parke,
  arXiv:1009.0014 [hep-ph].


\bibitem{ino}
  V.~Arumugam {\it et al.}  [INO Collaboration],
See, http://www.imsc.res.in/~ino/OpenReports/INOReport.pdf


\bibitem{Kato:2008zz}
  I.~Kato [ T2K Collaboration ],
  J.\ Phys.\ Conf.\ Ser.\  {\bf 136}, 022018 (2008).

\bibitem{Ayres:2004js}
  D.~S.~Ayres {\it et al.} [ NOvA Collaboration ],
   [hep-ex/0503053].








\bibitem{Kim:1998bv}
  C.~W.~Kim and U.~W.~Lee,
  Phys.\ Lett.\  B {\bf 444}, 204 (1998)
  [arXiv:hep-ph/9809491].



\bibitem{Peres:2003wd}
  O.~L.~G.~Peres and A.~Y.~Smirnov,
  Nucl.\ Phys.\  B {\bf 680}, 479 (2004)
  [arXiv:hep-ph/0309312].

\bibitem{Peres:2009xe}
  O.~L.~G.~Peres and A.~Y.~Smirnov,
  Phys.\ Rev.\  D {\bf 79}, 113002 (2009)
  [arXiv:0903.5323 [hep-ph]].

\bibitem{GonzalezGarcia:2004cu}
  M.~C.~Gonzalez-Garcia, M.~Maltoni and A.~Y.~Smirnov,
  Phys.\ Rev.\  D {\bf 70}, 093005 (2004)
  [arXiv:hep-ph/0408170].




\bibitem{Choubey:2005zy}
  S.~Choubey and P.~Roy,
  Phys.\ Rev.\  D {\bf 73}, 013006 (2006)
  [arXiv:hep-ph/0509197].

\bibitem{Indumathi:2006gr}
  D.~Indumathi, M.~V.~N.~Murthy, G.~Rajasekaran and N.~Sinha,
  Phys.\ Rev.\  D {\bf 74}, 053004 (2006)
  [arXiv:hep-ph/0603264].


\bibitem{Bandyopadhyay:2007kx}
  A.~Bandyopadhyay {\it et al.}  [ISS Physics Working Group],
  Rept.\ Prog.\ Phys.\  {\bf 72}, 106201 (2009)
  [arXiv:0710.4947 [hep-ph]].

\bibitem{Samanta:2008af}
  A.~Samanta,
  Phys.\ Rev.\  D {\bf 80}, 113003 (2009)
  [arXiv:0812.4639 [hep-ph]].

\bibitem{Akhmedov:2008qt}
  E.~K.~Akhmedov, M.~Maltoni and A.~Y.~Smirnov,
  JHEP {\bf 0806}, 072 (2008)
  [arXiv:0804.1466 [hep-ph]].


\bibitem{Dziewonski:1981xy}
  A.~M.~Dziewonski and D.~L.~Anderson,
  Phys.\ Earth Planet.\ Interiors {\bf 25}, 297 (1981).

\bibitem{Casper:2002sd}
  D.~Casper,
  Nucl.\ Phys.\ Proc.\ Suppl.\  {\bf 112}, 161 (2002)
  [arXiv:hep-ph/0208030].

\bibitem{geant}
http://geant4.web.cern.ch/geant4/




\bibitem{Samanta:2008ag}
  A.~Samanta,
  Phys.\ Rev.\ D {\bf 79}, 053011 (2009)
  [arXiv:0812.4640 [hep-ph]].

\bibitem{Honda:2006qj}
  M.~Honda, T.~Kajita, K.~Kasahara, S.~Midorikawa and T.~Sanuki,
  Phys.\ Rev.\  D {\bf 75}, 043006 (2007)
  [arXiv:astro-ph/0611418].


\bibitem{Samanta:2009qw}
  A.~Samanta,
  Phys.\ Rev.\  D {\bf 81}, 037302 (2010)
  [arXiv:0907.3540 [hep-ph]].

%












\bibitem{Aharmim:2009gd}
  B.~Aharmim {\it et al.}  [SNO Collaboration],
  Phys.\ Rev.\  C {\bf 81}, 055504 (2010)
  [arXiv:0910.2984 [nucl-ex]].

\bibitem{Mezzetto:2010zi}
  M.~Mezzetto and T.~Schwetz,
  J.\ Phys.\ G {\bf 37}, 103001 (2010)
  [arXiv:1003.5800 [hep-ph]].


\bibitem{Hiraide:2006vh}
 K.~Hiraide, H.~Minakata, T.~Nakaya, H.~Nunokawa, H.~Sugiyama,
W.~J.~C.~Teves, R.~Zukanovich Funchal,
neutrino oscillation experiments,''
 Phys.\ Rev.\  {\bf D73}, 093008 (2006)
 [hep-ph/0601258].

\bibitem{Kajita:2006bt}
 T.~Kajita, H.~Minakata, S.~Nakayama, H.~Nunokawa,
identical detectors with different baselines,''
 Phys.\ Rev.\  {\bf D75}, 013006 (2007)
 [hep-ph/0609286].

\bibitem{Meloni:2008it}
 D.~Meloni, O.~Mena, C.~Orme, S.~Palomares-Ruiz, S.~Pascoli,
 JHEP {\bf 0807}, 115 (2008)
 [arXiv:0802.0255 [hep-ph]].

\bibitem{Huber:2009cw}
  P.~Huber, M.~Lindner, T.~Schwetz, W.~Winter,
  JHEP {\bf 0911}, 044 (2009)
  [arXiv:0907.1896 [hep-ph]].


\bibitem{Huber:2004ug}
  P.~Huber, M.~Lindner, M.~Rolinec, T.~Schwetz and W.~Winter,
  Phys.\ Rev.\  D {\bf 70}, 073014 (2004)
  [arXiv:hep-ph/0403068].

\bibitem{Huber:2004dv}
  P.~Huber, M.~Lindner, M.~Rolinec, T.~Schwetz and W.~Winter,
  Nucl.\ Phys.\ Proc.\ Suppl.\  {\bf 145}, 190 (2005)
  [arXiv:hep-ph/0412133].





\end{thebibliography}
\end{document}